\documentclass[reprint, amsmath, amssymb, floatfix, aps, prx,nofootinbib,longbibliography]{revtex4-2}

\usepackage{bm}
\usepackage{graphicx}
\usepackage{hyperref}
\usepackage{bbold}
\usepackage{mathtools}
\usepackage[dvipsnames]{xcolor}
\usepackage{nicematrix}
\usepackage{natbib}
\usepackage{pgfplots}
\usepackage{tikz}

\hypersetup{
    colorlinks=true,
    linkcolor=blue,
    filecolor=magenta,      
    urlcolor=blue,
    pdftitle={Overleaf Example},
    pdfpagemode=FullScreen,
    }

% ...

\graphicspath{{./assets/figures/}}

\begin{document}

\preprint{}

\title{Multipartite Gaussian Entanglement Measure with Applications to Graph States and Bosonic Field Theory}

\author{Matteo Gori}
\email{matteo.gori@uni.lu}
\affiliation{
Department of Physics and Materials Science, University of Luxembourg, L-1511 Luxembourg City, Luxembourg
}

\author{Matthieu Sarkis}
\email{matthieu.sarkis@uni.lu}
\affiliation{
Department of Physics and Materials Science, University of Luxembourg, L-1511 Luxembourg City, Luxembourg
}

\author{Alexandre Tkatchenko}
\email{alexandre.tkatchenko@uni.lu}
\affiliation{
Department of Physics and Materials Science, University of Luxembourg, L-1511 Luxembourg City, Luxembourg
}

\date{\today}

\begin{abstract}
Computationally feasible multipartite entanglement measures are essential for advancing our understanding of complex quantum systems. Entanglement Distance (ED), introduced by \href{https://journals.aps.org/pra/abstract/10.1103/PhysRevA.101.042129}{Cocchiarella et al. \textit{Phys. Rev. A}, \textbf{101} (4), 042129 (2019)}, based on the Fubini-Study metric, offers several advantages over existing methods, including ease of computation, a profound geometrical interpretation, and applicability to multipartite entanglement. Although ED has been successfully applied to systems of qudits, an explicit formulation for continuous quantum variables, particularly for pure Gaussian states, remains unexplored. In this work, we address this limitation by deriving the analytical expression for the Gaussian Entanglement Measure (GEM), a multipartite entanglement monotone for multimode pure Gaussian states based on the purity of fragments of the whole systems, through a generalization of ED to group-theoretic coherent states. We show the efficacy of GEM across various scenarios, including the analysis of a two-mode Gaussian state under beam-splitter and squeezing transformations, and exploring graph states involving three and four modes. Notably, comparing GEM values for different graph topologies reveals insights into the connectivity of underlying graphs. Additionally, we illustrate how GEM provides insights into free bosonic field theory on $\mathbb R_t\times S^1$ beyond standard bipartite entanglement entropy, paving the way for quantum information-theoretical tools to probe the topological properties of quantum field theory spaces.
\end{abstract}

\keywords{Quantum Entanglement, Quantum Correlations}

\maketitle

    \begin{figure*}
    \includegraphics[scale=1]{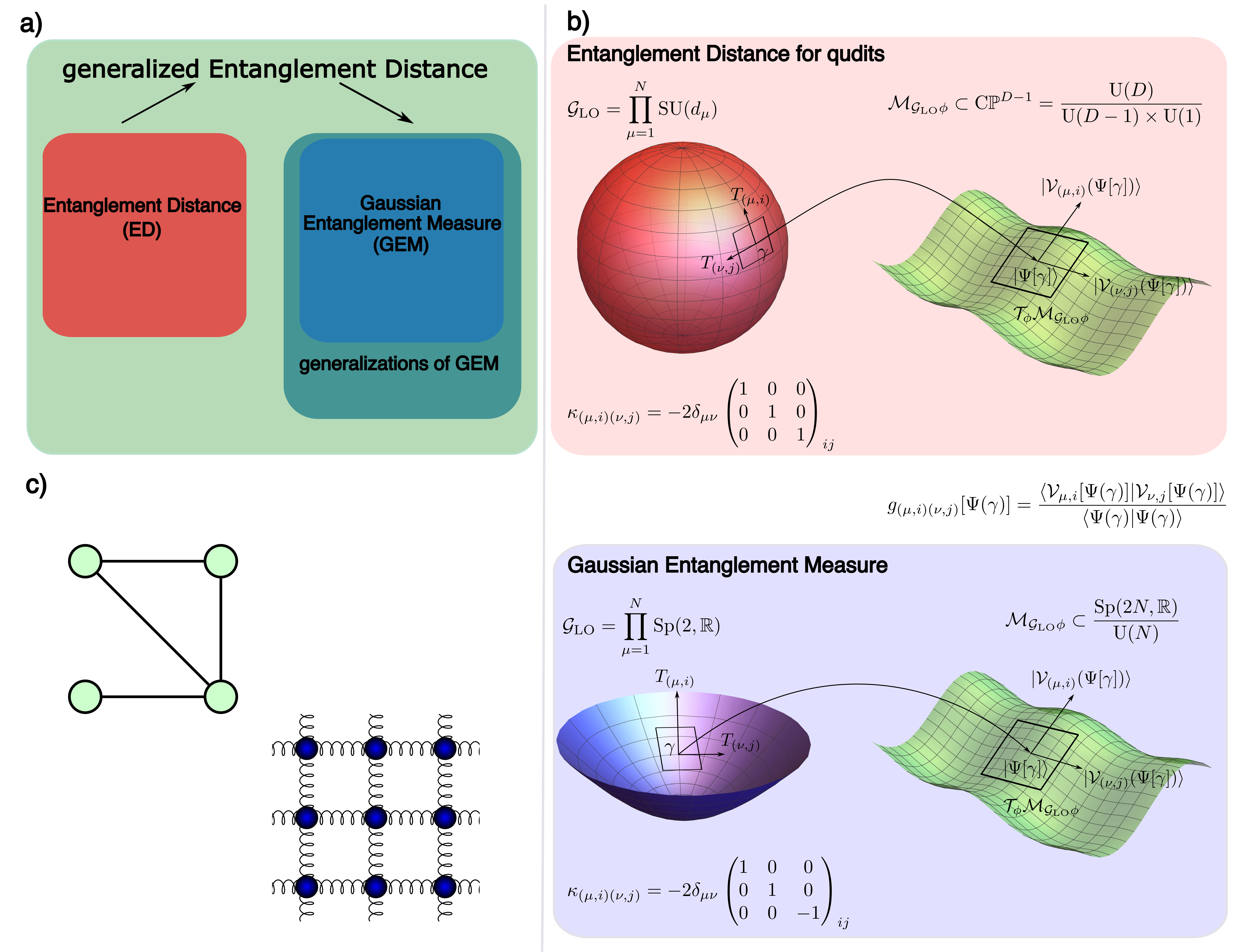}
    \caption{\textbf{Gaussian Entanglement Measure: concepts and applications} in panel a), illustration of the path followed in the paper to derive the Gaussian Entanglement Measure from the Entanglement Distance (ED) introduced in \cite{cocchiarella2020entanglement}: reformulation and generalization of the ED in the framework of group-theoretic coherent states, and definition of the GEM. In panel b), schematic illustration of the geometric construction involved in the paper. In the context of group-theoretic coherent states, we obtain the generalized ED of a given state $|\Psi\rangle$ by contracting the inverse of the Killing form $k_{(\mu,i)(\nu,j)}$ with the Fubini-Study metric $g_{(\mu,i)(\nu,j)}$ defined on the manifold $\mathcal{M}_{\mathcal{G}_{\rm LO}\Psi}$, which represents the orbit generated by the action of $\mathcal{G}_{LO}$ on the state $|\Psi\rangle$. The upper panel corresponds to a system of qudits, for which the group of local transformations $\mathcal{G}_{\rm LO}$ is compact. In that case, the Killing two-form is proportional to the identity, adopting the Gell-Mann normalization for $\rm{SU}(d_{\mu})$ generators $T_{(\mu,i)}$ of LOs for the $\mu$-th qudits. The lower panel corresponds to a system of harmonic oscillators (qumodes), which in the Gaussian case can be viewed as a non-compact analogue of the qudit system. In this case, the LO group is the symplectic group $\rm{Sp}(2,\mathbb{R})$, whose Killing two-form is not negative-definite.
    In panel c), possible applications of the GEM to the study of multipartite entanglement properties of graph states and discretized scalar fields.}
    \end{figure*}

    Quantum entanglement stands as a cornerstone in the 
    development of quantum-based technologies due to its 
    significance in the pursuit of quantum supremacy ~\cite{arute2019quantum,wu2021strong} and 
    enhanced quantum control ~\cite{ma2021quantum,tebbenjohanns2021quantum}. 
    Considerable strides have been made in comprehending quantum entanglement and its robustness within the mathematical framework of quantum information theory, encompassing both pure and mixed states~\cite{vidal1999robustness,vidal2000entanglement}.
    However, accurately quantifying entanglement continues to pose a significant unresolved challenge~\cite{vedral1997quantifying,horodecki2000limits,eisert2002quantification,plenio2007introduction,lancien2016should}.
    Historically, quantifying entanglement has centered on entropy-based measures. For instance, in pure and bipartite systems, the entropy of entanglement has been commonly acknowledged as a key measure of entanglement \cite{bennett1996mixed,popescu1997thermodynamics,vidal1999entanglement}. Faithful measures for mixed states of the same class of system 
    include 
    the entropy of formation~\cite{wootters1998entanglement},
    the entropy of distillation~\cite{bennett1996mixed}, and relative entropy of entanglement~\cite{vedral1997quantifying,bennett1996purification,horodecki1998mixed}. 
    Yet, extending these measures to encompass broader classes of quantum states, especially in multipartite systems, requires exploring diverse methodologies~\cite{shimony1995degree,dur2000three,briegel2001persistent,coffman2000distributed,yu2005multipartite}. Several measures, including the Schmidt measure~\cite{eisert2001schmidt} and generalized concurrence~\cite{carvalho2004decoherence}, have been proposed for multipartite systems, applicable to both pure and mixed states.
    The quest for a precisely defined entanglement measure capable of capturing multipartite entanglement across a wide range of systems beyond qubits, including both pure and mixed states, has prompted the investigation of novel approaches for deriving such measures.
    Recently, methods for estimating quantum entanglement and quantifying complexity in multipartite systems have emerged, leveraging insights from information geometry, particularly the quantum Fisher information~\cite{pezze2009entanglement,hyllus2012fisher,scali2019entanglement}.\\
    More recently, a multipartite entanglement 
    measure, called Entanglement Distance (ED)~\cite{cocchiarella2020entanglement,vesperini2023entanglement,vesperini2023unveiling, VESPERINI2023169406} has been proposed in the form of a 
    well-defined entanglement monotone derived from the Fubini-Study 
    metric defined on the projective Hilbert space of the pure quantum 
    states of a system of $M$ qudit states. Extensions of the ED have 
    been presented to include mixed states~\cite{vesperini2023entanglement}.\\
    The ED presents numerous advantages with respect to other multipartite
    entanglement monotones. From a conceptual point of view, it naturally provides a differential geometrical framework to describe 
    quantum entanglement~\cite{vesperini2023unveiling,VESPERINI2023169406}.
    From a computational point of view, the ED for a given pure state is \textit{a priori} much more efficient with respect to other multipartite 
    entanglement measures, as it does not require the optimization over a 
    set of separable states, but just the evaluation of a set of observables 
    (i.e., products of the generators of the group of local operations) on 
    the considered state. Applications of the ED have been presented for graph states~\cite{vesperini2023entanglement} and to
    the characterization of the entanglement protection of qubits in lossy cavities~\cite{nourmandipour2021entanglement}. 
    While, in principle, a similar construction could be done in the case of continuous variable quantum systems, no extensive studies have been
    done so far in this direction. 
    Gaussian states hold a pivotal position among the manifold of states linked to continuous quantum variables. They are known by various names (such as squeezed coherent states, generalized Slater determinants, and ground states of quadratic Hamiltonians) and find utility across diverse research domains, including quantum information~\cite{braunstein2005quantum,wang2007quantum,weedbrook2012gaussian,adesso2014continuous}, 
    in quantum field theory in curved spacetime~\cite{wald1994quantum,parker2009quantum} and in
    thermofield dynamics~\cite{umezawa1982thermo,blasone2011quantum}.
    In particular, Gaussian states constitute a
    manifold of states in (projective) Hilbert space 
    often used to approximate ground states for complicated 
    Hamiltonians through variational methods \cite{berazin2012method,guaita2019gaussian,windt2021local,cerezo2021variational}. Furthermore, numerous quantum information techniques can be analytically applied to Gaussian states, including computations of entanglement entropy \cite{chen2005gaussian,de2022linear}, logarithmic negativity \cite{adesso2005gaussian,adesso2006multipartite,tserkis2017quantifying}, and circuit complexity \cite{hackl2018circuit,chapman2019complexity}).
    Finally, recent studies in mathematical physics have
    revealed the rich differential geometrical structure of 
    Gaussian states (both for bosons and fermions), clearly 
    enlightening the connection with the theory of K\"ahler manifolds
    and generalized group theoretic coherent 
    states~\cite{hackl2020geometry,hackl2021bosonic}.\\
    This work expands on the concept of entanglement distance to explore multipartite entanglement in multimode bosonic Gaussian states. 
    Specifically, we leverage the geometric framework introduced in a previous study~\cite{hackl2020geometry} to derive the Gaussian Entanglement Measure (GEM), a multipartite entanglement monotone extending the concept of 
    ED to pure bosonic Gaussian state manifold.
    Such an extension is non trivial as it requires a generalization of the approach adopted in~\cite{cocchiarella2020entanglement} on building a scalar entanglement monotone from the Fubini-Study metric of the Hilbert space restricted  to the states that are locally invariant up to local operations
    (LOs). 
    Specifically, this involves extending the procedure for a set of
    qudits (where the LOs form a compact Lie group) to the case of 
    local Gaussian transformations acting on multimode bosonic
    Gaussian states, forming a non-compact semi-simple Lie group.
    For providing such an extension, we conveniently reformulate the ED in the language of group-theoretic
    coherent states and we clarify the close relation between the geometry of the Lie group of LOs and the geometry of the 
    submanifolds of states equivalent up to LOs.
    The GEM derived in the current work shares with the ED all the properties characterizing an entanglement measure. In particular, as we will see, the GEM admits a simple analytical expression for every
    pure gaussian state in terms of the purities of each part of the considered system. This confers a significant computational advantage, as computing the 
    GEM does not necessitate optimization across the set of separable Gaussian 
    states. Thanks to these features, the GEM is suitable tool that can be applied 
    to a wide range of systems. 
    Among the possible interesting applications of our 
    results,
    we treat the case of a bosonic scalar field in two spacetime dimensions. 
    The interplay between quantum information and quantum field theory is a very 
    active area of research, in particular in the framework of holography \cite{RevModPhys.90.035007, Calabrese_2009, Casini_2009, RevModPhys.90.045003, hollands2017entanglement, Rangamani:2016dms}.
    Traditionally, bipartite entanglement measures such as entanglement entropy 
    have been widely employed to explore the quantum entanglement of quantum 
    fields across two disjoint regions in spacetime.
    Additionally, our approach to the multipartite entanglement properties of Gaussian states could offer a new  
    quantum information theoretical tool for 
    exploring quantum phase transitions in 
    condensed matter systems.
    Recent research has highlighted the 
    significance of multipartite entanglement 
    measures linked to quantum Fisher 
    information~\cite{hyllus2012fisher,toth2012multipartite} (as the entanglement distance~\cite{cocchiarella2020entanglement})
    in gaining deeper insights into many-body 
    correlations~\cite{pezze2014quantum,hauke2016measuring},
    thus paving the way toward comprehension of quantum 
    phase transitions both in quantum spin systems and in 
    conformal field  theory \cite{rajabpour2017multipartite}.
    The paper is structured as follows. In Sec. \ref{sec:GEMonGCSM}
    we will review the definition of the ED introduced in~\cite{cocchiarella2020entanglement} and reformate it within the framework of 
    group-theoretic coherent state using the geometrical language
    presented in~\cite{hackl2020geometry}.
    We will clarify the introduction of the ED as the scalar invariant 
    obtained by taking the trace of the Fubini-Study metric on the submanifold of states 
    equivalent up to LOs as the contraction with the inverse of the Killing metric defined on the 
    algebra of the LOs group. This reformulation allows us in Sec.\ref{sec:GEM_definition}, in the context of continuous variable Gaussian states, to define and derive an analytical expression the Gaussian Entanglement Measure 
    (GEM). We discuss its general properties 
    and how it is related to other known multipartite entanglement measures for 
    Gaussian states. Finally, in Sec.\ref{sec:examples} we 
    present the application 
    of the GEM to what we will refer to as 'graph state', and to the ground state of a scalar 
    quantum field theory in two spacetime dimensions.
    The Conclusion section provides a 
    discussion of how the results obtained in the paper suggest several natural 
    generalizations, possible applications concerning the use of the GEM for 
    better understanding graph theoretic properties of multimode Gaussian states, 
    as well as for potentially being able to capture topological properties of the 
    spacetime on which a field theory is put.
    
    \paragraph*{Notations:} Lower case Greek letters denote the mode index. Lower case Latin letters denote Lie algebra indices. Upper case Latin letters denote coordinate indices in phase space.

\section{Entanglement distance from generalized coherent state manifold: general geometric formulation}
\label{sec:GEMonGCSM}

    \subsection{Fubini-Study metric on projective Hilbert space}
        In introducing the geometrical properties of the manifold of generalized 
        coherent state generated by the action of the ocal operation (LO) transformation group $\mathcal{G}_{\textsc{lo}}$, we will
        strictly follow \cite{hackl2020geometry}.\\
        Consider a finite quantum system $\mathcal{S}$ constituted by $N$ distinguishable physical elements. 
        The wave function for the isolated $i$-th part 
        belongs to the Hilbert space $\mathcal{H}_i$ so that all the wavefunction 
        $|\Psi \rangle $ the of the full system belongs to the 
        Hilbert space $\mathcal{H}=\bigotimes_{\mu=1}^{N} \mathcal{H}_\mu$.\\
        The quantum state of the total systems is represented by the projective
        Hilbert space 
        \begin{equation}
          \mathbb{P}(\mathcal{H}) = (\mathcal{H}\setminus \{\boldsymbol{0}\})/\sim\,,
        \end{equation}
        where two wavefunctions are equivalent if they are related through scaling by a non-zero complex number
        \begin{equation}
            |\Psi \rangle \sim |\Psi' \rangle \iff \exists\, c\in \mathbb{C}^*\ \ \text{such that}\ \ |\Psi\rangle =c |\Psi'\rangle\,\,.
        \end{equation}
        The tangent space to the projective Hilbert space at the point $|\Psi\rangle$ is defined as 
        \begin{equation}
            \mathcal{T}_{\Psi}\mathbb{P}(\mathcal{H})=\mathcal{H}/\sim\,,
        \end{equation}
        where the following equivalence relation has been introduced 
        \begin{equation}
            |X\rangle \approx |X'\rangle \iff \exists\, c\in \mathbb{C}^* \ \text{such that}\ |X'\rangle =c|X\rangle\,\,.
        \end{equation}
        From such a definition, it follows that the tangent space to a state vector
        $|\Psi\rangle$ in projective Hilbert space can be identified with the space of
        orthogonal vectors to $|\Psi\rangle$, i.e.
        \begin{equation}
            \mathcal{T}_{\Psi}\mathbb{P}(\mathcal{H})=\mathcal{H}^{\perp}_{\Psi}=\{|X \rangle \in\mathcal{H} \,\,\,|\,\,\, \langle X | \Psi \rangle =0\}\,\,.
        \end{equation}
        We notice that the tangent space to a state vector $|\Psi\rangle$ has the 
        structure of a Hilbert space induced by the Hilbert space structure on $\mathcal{H}$. In fact, a projector $\mathbb{Q}_{\Psi}$ acting on the space of state vector variations $|\delta \Psi \rangle$ at $|\Psi\rangle$  can be introduced such that
        \begin{equation}
        \label{eq:proj_PT}
            \mathbb{Q}_{\Psi}|\delta \Psi \rangle =   |\delta \Psi \rangle -\dfrac{\langle \Psi | \delta\Psi \rangle}{\langle \Psi|\Psi\rangle}|\Psi\rangle\,\,.
        \end{equation}
        In what follows, we will consider the case of a variational manifold $\mathcal{M}\subset \mathcal{H}$ of vector states $|\Psi(\boldsymbol{x})\rangle$ 
        parametrized by a set of real parameters $\boldsymbol{x}\in \mathcal{D}\subseteq\mathbb{R}^{m}$ (for some $m\in\mathbb N^*$), the tangent space $\mathcal{T}_{\boldsymbol{x}}\mathcal{M}$ at $|\Psi(\boldsymbol{x})\rangle$ is spanned by the vectors
        \begin{equation}
            |V_{A}(\boldsymbol{x})\rangle = \mathbb{Q}_{\Psi(\boldsymbol{x})}\dfrac{\partial}{\partial x^{A}}|\Psi(\boldsymbol{x})\rangle\,\,.
        \end{equation}
        forming a real basis $|V^{(\Psi)}_{A}\rangle$ of $\mathcal{T}_{\Psi}\mathbb{P}(\mathcal{H})$.
        Within this framework, it is possible to show that the inner
        product on the Hilbert space $\langle \cdot |\cdot \rangle$ induces
        a (real positive-definite) metric $g$ on $\mathcal{T}_{\Psi}\mathbb{P}(\mathcal{H})$, the well-known Fubini-Study metric, defined as
        \begin{equation}
            (g[\Psi])_{AB}=\dfrac{2\, \mathrm{Re}\left(\left\langle\left. V_{A}(|\Psi\rangle)\right|V_{B}(|\Psi\rangle) \right\rangle\right)}{\langle \Psi|\Psi\rangle}\,.
        \end{equation}
        The Fubini-Study metric endows the 
        variational manifold with a Riemannian structure.
        
        In \cite{cocchiarella2020entanglement,vesperini2023entanglement}
        a multipartite 
        entanglement measure for pure and mixed 
        states of a system 
        of qudits, the entanglement distance (ED), has been proposed based on 
        the Riemannian structure induced on the 
        projective Hilbert describing the system.
        In the next session, we will reformulate and broadly generalize this method within the 
        framework of generalized 
        coherent state manifold. This generalization will then allow us to define a new measure of multipartite entanglement associated to the submanifold of Gaussian states for
        a system of bosonic quantum harmonic oscillators (system of qumodes), the Gaussian Entanglement Measure (GEM).
    
    \subsection{Derivation of the entanglement distance from the geometry of group-theoretic coherent states}
    
 In this section, we rephrase the original derivation of 
 the entanglement distance (ED) for pure states, introduced 
 in~\cite{cocchiarella2020entanglement}, within the 
 geometrical framework of Gilmore-Perelmov group-theoretic coherent states. As extensively 
 discussed in~\cite{vesperini2023unveiling}, a key point at 
 the very basis of the ED for a state 
 $|\Psi\rangle$ is the definition of the equivalence class $[|\Psi\rangle]_{\rm LO}$ of states that are identical to $|\Psi\rangle$ up to a unitary transformation representing a local operation (LO). 
 If the ensemble of LOs forms a Lie Group $\mathcal{G}_{\textsc{lo}}$, the orbits $\mathcal{M}_{\mathcal{G}_{LO}\Psi}\subseteq [|\Psi\rangle]_{\rm LO}$ 
 generated by the application of unitary representations of 
 local operations $\mathcal{G}_{\textsc{lo}}$ to a state $|\Psi \rangle$ form a submanifold of group-theoretic coherent states, whose geometrical properties are well-known. The viewpoint we propose here clarifies the process of deriving the ED, which is a scalar quantity, obtained from the restriction of ambient Fubini-Study metric tensor of $\mathbb{P}(\mathcal{H})$ to $\mathcal{M}_{\mathcal{G}_{\mathrm{LO}}\Psi}$. Our approach provides a natural way to generalize the original ED derivation to more more complicated cases, for instance the situation where the LO group is the group of Gaussian transformations acting on Gaussian states.\\
By definition, an entanglement measure satisfies the 
condition~\cite{vidal1999entanglement} of being a non-
negative function defined on the set of equivalence 
classes $[|\Psi\rangle]_{\rm LO}$. 
In what follows, we assume that the considered set of local 
operations forms a real Lie group  $\mathcal{G}_{\textsc{lo}}$
admitting a real algebra $\mathfrak{g}_{\textsc{lo}}$ and a projective unitary representation $\mathcal{U}$ on $\mathcal{H}$, i.e.
\begin{equation}
        \mathcal{U}(\gamma_1) \mathcal{U}(\gamma_2) \sim \mathcal{U}(\gamma_1 \gamma_2)\,, \qquad \gamma_1,\gamma_2 \in \mathcal{G}_{\textsc{lo}}\,,
        \end{equation}
where the equivalence class $\sim$ identifies elements of the representation $\mathcal{U}$ that  differ for a complex phase factor.\\
For a composed system $\mathcal{S}$, we define for each part $\mu$ 
a Lie group action $\mathcal{G}_{\mu}$ with Lie algebra $\mathfrak{g}_{\mu}$,
such that the Lie algebra $\mathfrak{g}_{\textsc{lo}}$ of the considered local 
transformations is the direct sum of the Lie algebras of the Lie groups acting 
on the parts/fragments, i.e. $\mathfrak{g}_{\textsc{lo}}=\bigoplus_{\mu=1}^{M} \mathfrak{g}_{\mu}$. If $\{T_{(\mu,i)}\}_{i=1,...,n_{\mu}}$ is a basis for 
the algebra $\mathfrak{g}_{\mu}$, the product in the algebra $\mathfrak{g}_{\textsc{lo}}$ is defined by the brackets
        \begin{equation}
        \label{eq:commutation_rel}
        [T_{(\mu,i)},T_{(\nu,j)}]=\delta_{\mu \nu} [c^{(\mu)}]_{i j}^{k} T_{(\mu,k)}\,,
        \end{equation}
where $[c^{(\mu)}]_{i j}^{k} $ are the structure constants of the
algebra $\mathfrak{g}_{\mu}$. 
Recall at this point that a Lie group carries a natural invariant
symmetric bilinear form, the Killing form 
\begin{equation}
            \kappa_{(\mu, i) (\nu, j)}=\mathrm{Tr}(\mathrm{ad}_{T_{(\mu,i)}}\circ \mathrm{ad}_{T_{(\nu,j)}})=\delta_{\mu \nu}\left(\kappa^{(\mu)}\right)_{ij}\,.
\end{equation}
where $\left(\kappa^{(\mu)}\right)_{ij}=\left[c^{(\mu)}\right]_{i k}^{l} \left[c^{(\mu)}\right]_{j l}^{k}$  and $\mathrm{ad}_{T_{(\mu,i)}}\left(T_{(\nu,j)}\right)$ is the adjoint representation of the algebra
 $\mathcal{G}_{\textsc{lo}}$
 \begin{equation}
            \mathrm{ad}_{T_{(\mu,i)}}\left(T_{(\nu,j)}\right)=\left[T_{(\mu,i)}, T_{(\nu,j)}\right]=\delta_{\mu \nu} \left[c^{(\mu)}\right]_{ij}^{k}T_{(\mu,k)}\,.
 \end{equation}
A generic group element $g\in \mathcal{G}_{\textsc{lo}}$ can be expressed as  $\gamma=\prod_{\mu=1}^{N} \gamma_{\mu}$  where $\gamma_{\mu}\in\mathcal{G}_{\mu}$. 
We notice that the Lie group $\mathcal{G}_{\textsc{lo}}$ and, 
consequently, the associated multipartite entanglement 
measure will depend on how the system $\mathcal{S}$ is 
divided into $M$ parts, i.e., on the considered partition $\mathfrak{P}_{M}=\{\mathcal{S}_{1},...,\mathcal{S}_M\}$ of $\mathcal{S}$ with $\sqcup_{\mu=1}^{M} \mathcal{S}_{\mu} = \mathcal{S}$. 
In general, the number of parts $M$ is equal or smaller than the
number $N$ of physical elements of $\mathcal{S}$: however, 
we will focus here on finest (single-mode) partition i.e. $N=M$, and leave generalizations for future work \cite{sarkis}. \\
The action of the algebra of the Lie group on the Hilbert
space is defined as
\begin{equation}
        \hat{T}_{(\mu,i)}=\dfrac{\mathrm{d}}{\mathrm{d}s}\,\mathcal{U}\left(e^{i s T_{(\mu,i)}}\right)\biggr|_{s=0}\,.
\end{equation}
The local unitary operators are of the 
    form $\mathcal{U}(\gamma)=\prod_{\mu=1}^{N}\mathcal{U}^{(\mu)}(\gamma_{\mu})$
    where $\mathcal{U}^{(\mu)}(\gamma_\mu)$ acts non-trivially only on the
    Hilbert space $\mathcal{H}_{\mu}$.\\
%
%
 Given a representative state $|\phi\rangle$, one can generate the orbit 
under the (free) group action of $\mathcal{G}_{\textsc{lo}}$ on $\mathcal{M}$ 
generating the orbit  $\mathcal{M}_{\mathcal{G}_{\textsc{lo}}\phi}$, i.e. 
        \begin{equation}
            \mathcal{M}_{\mathcal{G}_{\textsc{lo}}\phi}=\{\mathcal{U}(\gamma)|\phi\rangle\ |\gamma \in\mathcal{G}_{\textsc{lo}}\}/\sim \ \subset \mathbb{P}(\mathcal{H})\,.
        \end{equation}
        A local system of coordinates on 
        $\mathcal{M}_{\mathcal{G}_{\textsc{lo}}\phi}$ in a neighbourhood of 
        $|\Psi(\gamma)\rangle = \mathcal{U}(\gamma)|\phi\rangle$ is induced by the exponential map
        \begin{equation}
            |\Psi(\gamma),\boldsymbol{s}\rangle = \mathcal{U}(\gamma)\mathcal{U}\left(e^{i\sum_{\mu=1}^N\sum_{i_{\mu}=1}^{n_{\mu}} s_{(\mu,i_{\mu})} T_{(\mu,i_{\mu})}}\right) |\phi \rangle \,.
        \end{equation}
        so that a basis of tangent vectors $|\mathcal{V}_{(\mu ,i )}(\gamma)\rangle = \mathbb{Q}_{\Psi(\gamma)}\mathcal{U}(\gamma)\hat{T}_{(\mu,i)}|\phi\rangle $ can be defined on the tangent space $\mathcal{T}_{\Psi(\gamma)} \mathcal{M}_{\mathcal{G}_{\textsc{lo}}\phi}$.
        The Fubini-Study metric on $\mathbb{P}(\mathcal{H})$
        induces a metric tensor $g$ on the manifold $\mathcal{M}_{\mathcal{G}_{\textsc{lo}}\phi}$ given by:
        \begin{equation}
        (g[\Psi(\gamma)])_{(\mu,i)(\nu,j)}  = \dfrac{2\, \mathrm{Re}\big\langle\left. \mathcal{V}_{(\mu,i)} (|\Psi(\gamma)\rangle)\right| \mathcal{V}_{(\nu,j)}(|\Psi(\gamma)\rangle)\big\rangle}{\langle \Psi(\gamma)|\Psi(\gamma)\rangle}\,,
        \end{equation}
        that has the property to be independent of $g\in\mathcal{G}_{\textsc{lo}}$  (see~\cite{hackl2020geometry,vesperini2023unveiling}), i.e.
        \begin{equation}
        \label{eq:metric_first}
            (g[\phi])_{(\mu,i)(\nu,j)}=-\dfrac{\langle \phi|\left(\hat{T}_{(\mu,i)}\mathbb{Q}_{\phi}\hat{T}_{(\nu,j)}+\hat{T}_{(\nu,j)}\mathbb{Q}_{\phi}\hat{T}_{(\mu,i)}\right)|\phi\rangle}{2\langle \phi|\phi\rangle}\,.
        \end{equation}
 Using the definition of the projector $\mathbb{Q}_{\phi}$ in Eq.\eqref{eq:proj_PT} we obtain that 
        \begin{equation}
        \label{eq:project_Tgen}
            \mathbb{Q}_{\phi} \hat{T}_{(\nu,j)}|\phi\rangle = \left(\hat{T}_{(\nu,j)} - \langle \hat{T}_{(\nu,j)}\rangle_{\phi} \right)|\phi\rangle = [\Delta_{\phi} \hat{T}_{(\nu,j)}] |\phi \rangle\,, 
        \end{equation}
        and substituting the previous expression into  
        \eqref{eq:metric_first} we obtain
        \begin{equation}
        \begin{aligned}
            &(g[\phi])_{(\mu,i)(\nu,j)}=\\ 
            &=-\dfrac{\left\langle[\Delta_{\phi} \hat{T}_{(\mu,i)}][\Delta_{\phi} \hat{T}_{(\nu,j)}]\right\rangle_{\phi}+(\mu,i) \leftrightarrow (\nu,j) }{2}\,.
        \end{aligned}
        \end{equation}
        It can be directly rewritten 
            \begin{equation}
        \label{eq:matric_homoman}
                (g[\phi])_{(\mu,i)(\nu,j)} = -\frac{M_{ij}^{\mu\nu}+M_{ji}^{\nu\mu}}{2} + M_i^\mu M_j^\nu\,.
            \end{equation}
            in terms of the first and second moments in state $|\phi\rangle$ of the $\mathfrak{g}_{\textsc{lo}}$ generators 
            \begin{equation}
            \label{eq:2nd_moments_gen}
                M_i^\mu = \left\langle \phi\left|T_{(\mu,i)}\right|\phi\right\rangle\,,\ \ \ \ M_{ij}^{\mu\nu} = \left\langle \phi\left|T_{(\mu,i)}T_{(\nu,j)}\right|\phi\right\rangle\,.
            \end{equation}
            It is worth noticing that the diagonal components of the
            Fubini-Study metric defined in eq. (\ref{eq:matric_homoman}) are proportional to the quantum Fisher
            information of the generator of the
            group $\hat{T}_{\mu,i}$ in the pure state $\rho= |\phi \rangle\langle \phi |$.\\
            As previously mentioned, in~\cite{cocchiarella2020entanglement} the specific case of a system of $N$ $d$-dimensional qudits is considered. Moreover, the authors consider the case of the finest partition for which $M=N$. This situation corresponds to the following choice of group of local transformation:
            \begin{equation}
                \mathcal{G}_{\textsc{lo}}=\prod_{\mu=1}^N \mathrm{SU}(d_\mu)\,.
            \end{equation}
            From the metric tensor in eq. \eqref{eq:matric_homoman}, the ED has been originally introduced in~\cite{cocchiarella2020entanglement} considering the
            trace of $g[\phi]$ along both the subsystems and the generator of LO indices.
            i.e.
            \begin{equation}
            \label{eq:original_ED}
            \mathrm{ED}[\phi]=\sum_{\mu,\nu=1}^{M}\sum_{i=1}^{n_{\mu}}\sum_{j=1}^{n_{\nu}} 
            \left[\delta^{(\mu,i)(\nu,j)}g_{(\mu,i)(\nu,j)}\right]+b\,\,
            \end{equation}
            where $b\in\mathbb{R}$ is a constant ensuring that for 
            a separable state $\phi_{\rm separable}$ the ED is 
            zero, i.e.
            $\mathrm{ED}[\phi_{\rm separable}]=0$.
            In the original definition of the ED reported in eq.\eqref{eq:original_ED}
            the indices over the LO algebra are contracted with a Kroenecker delta 
            that which does not carry any explicit information on the geometry of the Lie group  $\mathcal{G}_{\mathrm{LO}}$. It can be noticed that the Kroenecker delta happens to be proportional to the inverse of the Killing form defined
            on the Lie algebra $\mathfrak{g}_{\textsc{lo}}=\bigoplus_{\mu=1}^{M}\mathfrak{su}(d_{\mu})$
            according to the conventions adopted in~\cite{cocchiarella2020entanglement}.
            With this interpretation at hand, we propose the following generalized definition of the entanglement distance as follows for arbitrary Lie algebras $\mathfrak{g}_{\textsc{lo}}$
            \begin{equation}
            \label{eq:GEM_general}
            \text{ED}[\phi] = a \sum_{\mu,\nu=1}^{N} \sum_{i=1}^{n_{\mu}} \sum_{j=1}^{n_{\nu}}[ \kappa^{(\mu,i)(\nu,j)}(g[\phi])_{(\mu,i)(\nu,j)}]+b\,,
            \end{equation}
            where $a,b\in \mathbb{R}$ are constants 
            that can be fixed to ensure the correct
            normalization of the entanglement measure;
            in particular, this generalized $\text{ED}$ is non-negative and 
            vanishes for separable states.
            The generalized ED defined in eq. \eqref{eq:GEM_general}
            reduces to the one in eq. \eqref{eq:original_ED} 
            for system of qudits in the particular case of $\mathfrak{g}_{\mu}=\mathfrak{su}(n_{\mu})$ 
            for which the Killing form is negative definite.
            
            However, the definition \eqref{eq:GEM_general} is more generic, and  differences with the compact special unitary case relevant for qudits start emerging when the 
            Killing form is not negative definite, i.e. for semi-simple non-compact Lie algebras as in the case of local Gaussian operations acting on a single bosonic mode where the local algebra is given by the real symplectic algebra $\mathfrak{g}_{\mu}=\mathfrak{sp}(2,\mathbb{R})$.
            In the next section, we will precisely apply the generalized ED definition presented in eq. \eqref{eq:GEM_general} to multimode bosonic Gaussian states.

\section{Gaussian Entanglement Measure for pure gaussian states}
\label{sec:GEM_definition}

    As described in the previous section, by specializing the generalized ED to the case of the special unitary algebra, say for instance $\mathfrak{g}_{\mu}=\mathfrak{su}(2)$, one recovers the ED  of a system of qubits. Instead, let us consider the case of a system of bosonic quantum harmonic oscillators (qumodes). The role of the local algebra is then played by the (non-compact) real symplectic algebra $\mathfrak{g}_{\mu}=\mathfrak{sp}(2, \mathbb R)$. The projectivized Hilbert space $\mathbb{CP}^1=\text{SU}(2)/\text{U}(1)$ (the Bloch sphere) for a single qubit is indeed  naturally replaced by the hyperboloid $\mathbb H^1=\text{Sp}(2,\mathbb R)/\text{U}(1)$ for a single qumode.

    \subsection{The geometry of Gaussian states manifold}
        Let us consider an $N$-modes bosonic 
        system defined by the algebra of creation
        annihilation operators
        \begin{equation}
        \mathcal{A}_{N}=\{\hat{a}_1,\hat{a}^{\dagger}_{1},...,\hat{a}_{N},\hat{a}^{\dagger}_{N}\}
        \end{equation}
        satisfying the canonical commutation relation:
        \begin{equation}
            [a_{\mu}, a_{\nu}^\dagger] = \delta_{\mu \nu}\,,
        \end{equation}
        and the space of pure states belonging to the (projective) 
        Hilbert space 
        $\mathbb{P}(\mathcal{H})$.\\
        It is convenient to introduce the algebra of 
        quadrature operators
        \begin{equation}
        \label{eq:xivar_def}
        \{\hat{q}_1,\hat{p}_1,...,\hat{q}_N,\hat{p}_N\}=\{\hat{\xi}_{1},...,\hat{\xi}_{2N}\}  \,,
        \end{equation}
        that are related to the creation/annihilation 
        operators by the transformation
        \begin{equation}
        \hat{a}_{\mu} = \frac{\hat{q}_{\mu} + i\hat{p}_{\mu}}{\sqrt 2}\,,\ \ \ \ \hat{a}_{\nu}^\dagger = \frac{\hat{q}_{\nu} - i\hat{p}_{\nu}}{\sqrt 2}\,.
        \end{equation}
        and satisfying the commutation relations
        \begin{equation}        
        [\hat{q}_{\mu}, \hat{p}_{\nu}] = i \delta_{\mu\nu}\,.
        \end{equation}
        
        For each state $[|\Psi\rangle] \in \mathbb{P}(\mathcal{H})$, 
        we define the $k$-point correlation coefficients 
        \begin{equation}
            \mathcal{C}_{A_1,\dots ,A_k}([|\Psi\rangle])= \langle(\hat{\xi}_{A_1}-\langle \hat{\xi}_{A_1} \rangle_{\Psi})\dots(\hat{\xi}_{A_k}-\langle \hat{\xi}_{A_k} \rangle_{\Psi})\rangle_{\Psi}\,\,.
        \end{equation}
        where the capital Latin letters 
        runs over the $2N$-dimensional
        phase space, i.e. $A_1,...,A_k=1,...,2N$.
        In particular, the 2-point correlation coefficients can be rewritten in terms of the matrices
        \begin{equation}
        \label{eq:corr_decomposition}
        \mathcal{C}_{A,B}=\Gamma_{AB}+\dfrac{i}{2}\,\Omega_{AB}\,,
        \end{equation}
        where we have introduced the
        \begin{equation}
        \label{eq:gamma_def}
            \Gamma_{AB} = \dfrac{1}{2}\left\langle \Psi \left|  \left\{ \hat{\xi}_{A},\hat{\xi}_{B}  \right\} \right| \Psi \right\rangle = \dfrac{\mathcal{C}_{AB}+\mathcal{C}_{BA}}{2}\,,
        \end{equation}
        and anti-symmetric (symplectic) matrix determined by the CCR
        \begin{equation}
        \label{eq:omega_def}
            \Omega_{AB}= -i \left\langle \Psi \left| \left[\hat{\xi}_{A},\hat{\xi}_{B}\right] \right|\Psi\right\rangle= -i(\mathcal{C}_{AB}-\mathcal{C}_{BA})\,,
        \end{equation}
        that, according to our conventions, can be rewritten as
        \begin{equation}
        \label{eq:Omegaform}
        \Omega=\bigoplus_{\mu=1}^N \Omega^{(\mu)}=\bigoplus_{\mu=1}^N
            \begin{pmatrix}
            0 & 1 \\
            -1  & 0 \\
            \end{pmatrix}\,.
        \end{equation}
        
        In what follows, we will consider the set 
        of Gaussian states parametrized by the matrix $\Gamma$ such that 
        \begin{equation}
        \label{eq:Gauss_def}
            \langle \xi_{A} \rangle_{\Gamma}=0 \,\,\,\text{for}\,\, A=1,...,2N \quad  \text{and} \quad \left(\Gamma\Omega^{-1}\right)^2=-\dfrac{\mathbb{1}}{4}\,\,.
        \end{equation}
        A property of the Gaussian states as defined in 
        eq.\eqref{eq:Gauss_def} is that all the $2n+1$-point
        correlation coefficients vanish while all the $2n$-point
        correlation coefficients can be expressed in terms of
        products of the $\Gamma$ matrix.\\
        We indicate the generic Gaussian state as 
        $|\Gamma\rangle$, and 
        whenever unambiguous,
        we will denote by simple brackets the expectation
        in state $|\Gamma\rangle$.\\
        A remarkable property of the Gaussian 
        states is the well-known Wick theorem:
        the $2n$-point functions in state $|\Gamma\rangle$ can be decomposed into the sum of products of 2-point functions
        \begin{equation}
        \label{eq:Wick_Thr}
            \mathcal C_{A_1,\dots,A_{2n}}(\Gamma) = \frac{1}{n!}\sum_\sigma\mathcal C_{A_{\sigma(1)},A_{\sigma(2)}}\dots\mathcal C_{A_{\sigma(2n-1)},A_{\sigma(2n)}}\,,
        \end{equation}
        where the sum is over all the permutations of $2n$ elements such that $\sigma(2i-1)<\sigma(2i)$ for all $i$.
        
    \subsection{Generators of Gaussian unitary transformations: single mode case}
        To estimate the multipartite entanglement 
        distance eq.\eqref{eq:GEM_general}
        for a generic multimode bosonic 
        Gaussian state, we need
        to define the generators of unitary representation of 
        local operations that are automorphisms
        in the space of bosonic Gaussian states. 
        We assume identifying each part of
        our system with a single bosonic mode.
        A group of automorphism admitting unitary 
        representation on the Gaussian 
        state manifold is the group of Gaussian unitary 
        operator generated by a quadratic Hamiltonian 
        of the form
        \begin{equation}
            \hat{H} = \frac{1}{2}\begin{pmatrix}
            \hat{a}^\dagger & \hat{a}
            \end{pmatrix}\begin{pmatrix}
                A & B \\ B^* & A
            \end{pmatrix}\begin{pmatrix}
                \hat{a} \\ \hat{a}^\dagger
            \end{pmatrix}\,,
        \end{equation}
        with the constraint that $A\in\mathbb R$ and $B \in \mathbb{C}$.
        One can expand it as
        \begin{equation}
            \hat{H} = 2\text{Re}(B)\,\hat{T}_1 - 2\text{Im}(B)\,\hat{T}_2 + 2A\,\hat{T}_3= \bm n\cdot  \hat{\bm T}
        \end{equation}
        where we have defined the generators 
        \begin{equation}
        \label{eq:generators_sp2}
        \begin{alignedat}{2}
            \hat{T}_1 &= \frac{\hat{a}^2 + (\hat{a}^\dagger)^2}{4} &&= \frac{\hat{q}^2 - \hat{p}^2}{4}\,, \\ 
                \hat{T}_2 &= -\frac{\hat{a}^2 - (\hat{a}^\dagger)^2}{4i} &&=- \frac{\hat{p}\hat{q} + \hat{q}\hat{p}}{4}\,, \\ 
                \hat{T}_3 &= \frac{\hat{a}^\dagger \hat{a} + \hat{a} \hat{a}^\dagger}{4} &&= \frac{\hat{q}^2 + \hat{p}^2}{4}\,,
            \end{alignedat}
        \end{equation}
        and the vector
        \begin{equation}
            \bm n = 2\left(
                \text{Re}(B), -\text{Im}(B), A
            \right)^\textsc{t}\,.
        \end{equation}
        One can check that the generators satisfy the $\mathfrak{sp}(2,\mathbb R)$ algebra
        \begin{align}
            [\hat{T}_1, \hat{T}_2] &= -i\hat{T}_3\,,\nonumber\\ [\hat{T}_2, \hat{T}_3] &= i\hat{T}_1\,,\\ [\hat{T}_3, \hat{T}_1] &= i\hat{T}_2\,.\nonumber
        \end{align}
        The generators are all Hermitian, and form an infinite dimensional representation of the symplectic algebra $\mathfrak{sp}(2,\mathbb R)$, known as the metaplectic representation \cite{wolf2013integral, arvind1995real}.
        The associated Killing form corresponds to the Lorentzian metric
        \begin{equation}
            \kappa_{ij} = 2\begin{pmatrix}
                -1 & 0  & 0 \\
                0  & -1 & 0 \\
                0  & 0  & 1
            \end{pmatrix}\,.
        \end{equation}
        We denote the inverse Killing form by $\kappa^{ij}$.

    \subsection{Exact computation of the metric tensor: derivation of the GEM}
    \label{sec:derivation_metric}
        To apply the general definition of the 
        GEM in eq.\eqref{eq:GEM_general} to
        a multimode pure Gaussian state, it is required
        to explicitly compute the (non-trivial) metric
        tensor defined on the Lie subgroup of single-mode 
        local operations discussed above.\\
        This necessitates the estimation of the expectation values 
        expressed in Eq .\eqref{eq:2nd_moments_gen} for both the 
        generators outlined in eq.\eqref{eq:generators_sp2} and their 
        pairwise products, i.e. 2-points and 4-points functions of the $\hat{q}_{\mu},\hat{p}_{\mu}$ that can be easily calculated by 
        Wick theorem in eq. (\ref{eq:Wick_Thr}).
        In the particular case of 4-point functions relevant to us, Wick's theorem gives:
            \begin{align}
                \mathcal C_{A_1,A_2,A_3,A_4} &=\mathcal C_{A_1,A_2}\mathcal C_{A_3,A_4} + \mathcal C_{A_1,A_3}\mathcal C_{A_2,A_4}\nonumber\\
                &+ \mathcal C_{A_1,A_4}\mathcal C_{A_2,A_3}\,,
            \end{align}
        while the two-point function can be expressed in terms of the correlation matrix and the symplectic form as in eq.~\eqref{eq:corr_decomposition}
        The metric tensor is then obtained by symmetrizing the connected component of the second moments. The reader will find the explicit expression of these moments in the App. (\ref{app:moments}). After the dust settles down, we obtain the following components of the metric tensor:
        \begin{widetext}
        \begin{equation}
        \begin{aligned}
            (g[\Gamma])_{(\mu,1)(\nu,1)} &= \frac{1}{8} \left(-\Gamma _{p_{\mu },p_{\nu }}^2+\Gamma _{p_{\mu },q_{\nu }}^2+\Gamma _{p_{\nu },q_{\mu }}^2-\Gamma _{q_{\mu },q_{\nu }}^2\right)-\frac{\delta _{\mu ,\nu }}{16}\,,\\
            (g[\Gamma])_{(\mu,1)(\nu,2)} &= \frac{1}{4} \left(\Gamma _{q_{\mu },q_{\nu }} \Gamma _{p_{\nu },q_{\mu }}-\Gamma _{p_{\mu },p_{\nu }} \Gamma _{p_{\mu },q_{\nu }}\right)\,,\\
            (g[\Gamma])_{(\mu,1)(\nu,3)} &= \frac{1}{8} \left(\Gamma _{p_{\mu },p_{\nu }}^2+\Gamma _{p_{\mu },q_{\nu }}^2-\Gamma _{p_{\nu },q_{\mu }}^2-\Gamma _{q_{\mu },q_{\nu }}^2\right)\,,\\
            (g[\Gamma])_{(\mu,2)(\nu,2)} &= \frac{1}{16} \left(-\delta _{\mu ,\nu }-4 \left(\Gamma _{p_{\mu },q_{\nu }} \Gamma _{p_{\nu },q_{\mu }}+\Gamma _{p_{\mu },p_{\nu }} \Gamma _{q_{\mu },q_{\nu }}\right)\right)\,,\\
            (g[\Gamma])_{(\mu,1)(\nu,3)} &= \frac{1}{4} \left(\Gamma _{p_{\mu },p_{\nu }} \Gamma _{p_{\nu },q_{\mu }}+\Gamma _{q_{\mu },q_{\nu }} \Gamma _{p_{\mu },q_{\nu }}\right)\,,\\
            (g[\Gamma])_{(\mu,3)(\nu,3)} &= \frac{1}{16} \left(\delta _{\mu ,\nu }-2 \left(\Gamma _{p_{\mu },p_{\nu }}^2+\Gamma _{p_{\mu },q_{\nu }}^2+\Gamma _{p_{\nu },q_{\mu }}^2+\Gamma _{q_{\mu },q_{\nu }}^2\right)\right)\,.
        \end{aligned}
        \end{equation}
        \end{widetext}
        In particular, contracting with the Killing form, we have:
        \begin{align}
            \text{GEM}[\Gamma]&=a\sum_{i,j,\mu,\nu}\kappa^{(\mu,i)(\nu,j)}(g[\Gamma])_{(\mu,i)(\nu,j)}+b\nonumber\\
        &=a\sum_{i,j,\mu}\kappa^{ij}(g[\Gamma])_{(\mu,i)(\mu,j)}+b\nonumber\\
            &=\frac{a}{8} \sum_{\mu=1}^{N}\left[\text{det}\left(\Gamma^{(\mu)}\right) -\frac{3}{4}\right]+b\,,
        \end{align}
        where the reduced density correlation 
        matrix is defined as
        \begin{equation}
            \Gamma^{(\mu)}= 
            \begin{pmatrix}
            \Gamma_{q_{\mu},q_{\mu}} & \Gamma_{q_{\mu},p_{\mu}}\\
            \Gamma_{p_{\mu},q_{\mu}} & \Gamma_{p_{\mu},p_{\mu}}
            \end{pmatrix}\,.
        \end{equation}
        
        To fix the normalization parameters $a$ and $b$, we note that for pure single-mode Gaussian states the eq.\eqref{eq:Gauss_def}
        reads
        \begin{equation}
        \label{eq:Gauss_pure_cond}
        \Gamma^{(\mu)}\left[\Omega^{(\mu)}\right]^{-1}=J^{(\mu)}\quad \text{and} \quad \left[J^{(\mu)}\right]^2=-\dfrac{\mathbb{1}}{4}\,.
        \end{equation}
        From the previous definitions, it follows that
        \begin{widetext}
        \begin{equation}
        \begin{aligned}
            &(\Gamma^{(\mu)}\left[\Omega^{(\mu)}\right]^{-1})^2=\left[\begin{array}{cc}
                (\mathcal C_{q_{\mu},p_{\mu}}+\mathcal C_{p_{\mu},q_{\mu}})^2-4 \mathcal C_{q_{\mu},q_{\mu}}\mathcal C_{p_{\mu},p_{\mu}} & 0\\
                0 & (\mathcal C_{q_{\mu},p_{\mu}}+\mathcal C_{p_{\mu},q_{\mu}})^2-4 \mathcal C_{q_{\mu},q_{\mu}}\mathcal C_{p_{\mu},p_{\mu}}\\
            \end{array}\right]\overset{!}{=}-\mathbb{1}\,,
        \end{aligned}
        \end{equation}
        \end{widetext}
        from which the following conditions for 2-point functions follow
        \begin{equation}
            4 \mathcal C_{q_{\mu},q_{\mu}}\mathcal C_{p_{\mu},p_{\mu}} - (\mathcal C_{q_{\mu},p_{\mu}}+\mathcal C_{p_{\mu},q_{\mu}})^2 = 1\,,
        \end{equation}
        or, equivalently, in terms of the correlation matrix:
        \begin{equation}
            \Gamma_{p_{\mu},q_{\mu}}^2 = \Gamma_{q_{\mu},q_{\mu}}\Gamma_{p_{\mu},p_{\mu}} - \frac{1}{4}\,.
        \end{equation}
        Therefore, one has for a separable state (setting here $a=1$ and $b=0$)
        \begin{equation}
            \text{GEM}[\Gamma_\text{separable}]=\frac{N}{8}\,.
        \end{equation}
        Therefore, setting $a=1$ and $b=-\text{GEM}[\Gamma_\text{separable}; 1,0]$,
        the Gaussian Entanglement Measure of a pure $M$-mode Gaussian state is then given by:
        \begin{equation}
            \text{GEM}[\Gamma] = \frac{1}{8}\sum_{\mu=1}^{N}\left[\text{det}\left(\Gamma^{(\mu)}\right) -\frac{1}{4}\right]\,.
        \end{equation}
        Let us recall that the purity of a quantum state described by the density matrix $\rho$ is defined as $P\left(\rho\right) = \text{tr}\left(\rho^2\right)$. For a Gaussian state identified by the covariance matrix $\Gamma$, the purity can be simply expressed \cite{paris2003purity} as:
        \begin{equation}
            P\left(\rho\right) = \frac{1}{2\sqrt{\text{det}\left(\Gamma\right)}}\,.
        \end{equation}
        Hence, the GEM can be rewritten in terms of the purities of the subsystems:
        \begin{equation}
        \label{eq:final_expression_GEM}
            \text{GEM}[\Gamma] = \frac{1}{32}\sum_{\mu=1}^{N}\left[\frac{1}{P\left(\rho^{(\mu)}\right)^2} - 1\right]\,.
        \end{equation}
        It can be very interestingly noticed that another quantity known as the 'potential of multipartite entanglement' $\pi_{\text{ME}}$ \cite{facchi2008maximally, facchi2009gaussian}, appears to be defined as an average of the purity over partitions of a quantum state into subsystems, $\pi_{\text{ME}}\propto \sum_\mu P(\rho^{(\mu)})$. Our approach can be viewed as providing a first-principles motivation for introducing the average purity of the subsystems. 
        Note that by linearity, one can subtract away the contribution of separable states already at the level of the full metric tensor by defining $h[\Gamma]= g[\Gamma]-g[\Gamma_\text{separable}]$ with the following non-zero components:    
        \begin{widetext}
        \begin{equation}
        \begin{aligned}
        \label{eq:metric_difference}
            (h[\Gamma])_{(\mu,1)(\nu,1)} &= \frac{1}{8} \left( -\Gamma _{p_{\mu },p_{\nu }}^2+\Gamma _{p_{\mu },q_{\nu }}^2+\Gamma _{p_{\nu },q_{\mu }}^2-\Gamma _{q_{\mu },q_{\nu }}^2+\left(\Gamma _{p_{\mu },p_{\mu }}-\Gamma _{q_{\mu },q_{\mu }}\right){}^2+1-\frac{\delta _{\mu ,\nu }}{2}\right)\,,\\
            (h[\Gamma])_{(\mu,1)(\nu,2)} &= \frac{1}{4} \left(-\Gamma _{p_{\mu },p_{\nu }} \Gamma _{p_{\mu },q_{\nu }}+\Gamma _{q_{\mu },q_{\nu }} \Gamma _{p_{\nu },q_{\mu }}+\Gamma _{p_{\mu },q_{\mu }} \left(\Gamma _{p_{\mu },p_{\mu }}-\Gamma _{q_{\mu },q_{\mu }}\right)\right)\,,\\
            (h[\Gamma])_{(\mu,1)(\nu,3)} &= \frac{1}{8} \left(\Gamma _{p_{\mu },p_{\nu }}^2-\Gamma _{p_{\mu },p_{\mu }}^2+\Gamma _{p_{\mu },q_{\nu }}^2-\Gamma _{p_{\nu },q_{\mu }}^2-\Gamma _{q_{\mu },q_{\nu }}^2+\Gamma _{q_{\mu },q_{\mu }}^2\right)\,,\\
            (h[\Gamma])_{(\mu,2)(\nu,2)} &= \frac{1}{4} \left(- \Gamma _{p_{\mu },q_{\nu }} \Gamma _{p_{\nu },q_{\mu }}- \Gamma _{p_{\mu },p_{\nu }} \Gamma _{q_{\mu },q_{\nu }}+2 \Gamma _{p_{\mu },p_{\mu }} \Gamma _{q_{\mu },q_{\mu }}-\frac{\delta _{\mu ,\nu }}{4}\right)\,,\\
            (h[\Gamma])_{(\mu,2)(\nu,3)} &= \frac{1}{4} \left(\Gamma _{p_{\mu },p_{\nu }} \Gamma _{p_{\nu },q_{\mu }}+\Gamma _{q_{\mu },q_{\nu }} \Gamma _{p_{\mu },q_{\nu }}-\Gamma _{p_{\mu },q_{\mu }} \left(\Gamma _{p_{\mu },p_{\mu }}+\Gamma _{q_{\mu },q_{\mu }}\right)\right)\,,\\
            (h[\Gamma])_{(\mu,3)(\nu,3)} &= \frac{1}{16} \left(- \Gamma _{p_{\mu },p_{\nu }}^2-\Gamma _{p_{\mu },q_{\nu }}^2-\Gamma _{p_{\nu },q_{\mu }}^2- \Gamma _{q_{\mu },q_{\nu }}^2+ \left(\Gamma _{p_{\mu },p_{\mu }}+\Gamma _{q_{\mu },q_{\mu }}\right){}^2-1+\frac{\delta _{\mu ,\nu }}{2}\right)\,.
        \end{aligned}
        \end{equation}
        \end{widetext}
        The GEM is then simply given by contraction with the inverse of the Killing metric as before:
        \begin{equation}
        \begin{aligned}
            \text{GEM}[\Gamma]&=\sum_{i,j,\mu,\nu}\kappa^{(\mu,i)(\nu,j)}(h[\Gamma])_{(\mu,i)(\nu,j)}\\
            &=\frac{1}{32}\sum_{\mu=1}^{N}\left[\frac{1}{P\left(\rho^{(\mu)}\right)^2} - 1\right]\,.
        \end{aligned}
        \end{equation}
        Let us mention at this point that we could actually choose to normalize the GEM slightly differently, in particular by dividing by a global factor of $N$, giving the GEM the interpretation of an arithmetic average over the subsystems. We do not choose such a normalization here, but we refer the reader to the end of Sec. \ref{subsec:scalar_field} for another comment about this other possible choice of normalization.

    \subsection{Properties satisfied by GEM}
    
        Generic entanglement measures are required to satisfy axioms \cite{PhysRevLett.78.2275, cocchiarella2020entanglement, RevModPhys.81.865, andersson2014quantum}, that we check below for our GEM measure.
        
        \paragraph*{Invariance under local unitaries:} This property is natural given the construction. By definition, our quantity only depends on the state up to single-mode unitary transformations. 
        
        \paragraph*{Positivity:} This property is obvious from the expression (\ref{eq:final_expression_GEM}) in terms of the purity.
        
        \paragraph*{Upper bound:} Due to the non-compactness of $\text{Sp}(2M,\mathbb R)$, one should not expect our definition to admit an upper bound. We refer the reader to Sec. \ref{sec:two-mode} below for a comment concerning a possible way to modify the GEM to make it upper-bounded.
        
        \paragraph*{Upper bound attained by maximally entangled states:} The states playing the role of the maximally entangled Bell states correspond in the continuous variable setting to non-normalizable states of the form $\sum_n|n\rangle\otimes\dots\otimes|n\rangle$, for which will see in Sec. \ref{sec:two-mode} that indeed their GEM diverges.
        
        \paragraph*{Vanishes on separable states:} By construction of the metric tensor $h$, eq. (\ref{eq:metric_difference}), the GEM attains its lower bound for separable states.

\section{Examples}
\label{sec:examples}
    Generic families of multi-mode Gaussian states are obtained using Hermitian Hamiltonians that are quadratic in the quadratures, or in the creation and annihilation operators. If we define multi-mode generators generalizing their single-mode counterpart (\ref{eq:generators_sp2}) as follows:
    \begin{equation}
        \begin{aligned}
        \hat{T}_{1,\mu\nu} &= \frac{\hat{a}_\mu \hat{a}_\nu + \hat{a}_\mu^\dagger \hat{a}_\nu^\dagger}{4}\,, \ \ \hat{T}_{2,\mu\nu} = -\frac{\hat{a}_\mu \hat{a}_\nu - \hat{a}_\mu^\dagger \hat{a}_\nu^\dagger}{4i}\,, \\
        \hat{T}_{3,\mu\nu} &= \frac{\hat{a}_\mu^\dagger \hat{a}_\nu + \hat{a}_\mu \hat{a}_\nu^\dagger}{4}\,, \ \ \hat{T}_{4,\mu\nu} = -\frac{\hat{a}_\mu^\dagger \hat{a}_\nu - \hat{a}_\mu \hat{a}_\nu^\dagger}{4i}\,,
    \end{aligned}
    \end{equation}
    one can indeed define the following families of $M$-mode Gaussian states:
    \begin{equation}
        \left|\left\{c_{i,\mu\nu}\right\}\right\rangle = U\left(\left\{c_{i,\mu\nu}\right\}\right)|0\rangle = e^{i\hat{H}\left(\left\{c_{i,\mu\nu}\right\}\right)}|0\rangle
    \end{equation}
    with the following generic Hermitian generator:
    \begin{equation}
    \label{eq:generic_preparation_hamiltonian}
        \hat{H}\left(\left\{c_{i,\mu\nu}\right\}\right) = \sum_{\mu,\nu=1}^{N}\sum_{i=1}^4 c_{i,\mu\nu}\hat{T}_{i,\mu\nu}\,.
    \end{equation}
    and real coefficients $c_{i,\mu\nu}$. For a given number of modes $N$, this family of states is $N(2N+1)$-dimensional, which is of course the dimension of the symplectic algebra $\mathfrak{sp}(2N,\mathbb R)$. 

    For visualization, we will illustrate the GEM for low dimensional sub-families of states which we will refer to as 'graph states' in what follows. 
    
    We also study the case of a massive Klein-Gordon field in two spacetime dimensions to illustrate the applicability of our results to systems with a large number of degrees of freedom. 

    \subsection{Graph states}
    
        \begin{figure*}[ht]
            \centering
            \includegraphics[scale = 0.7]{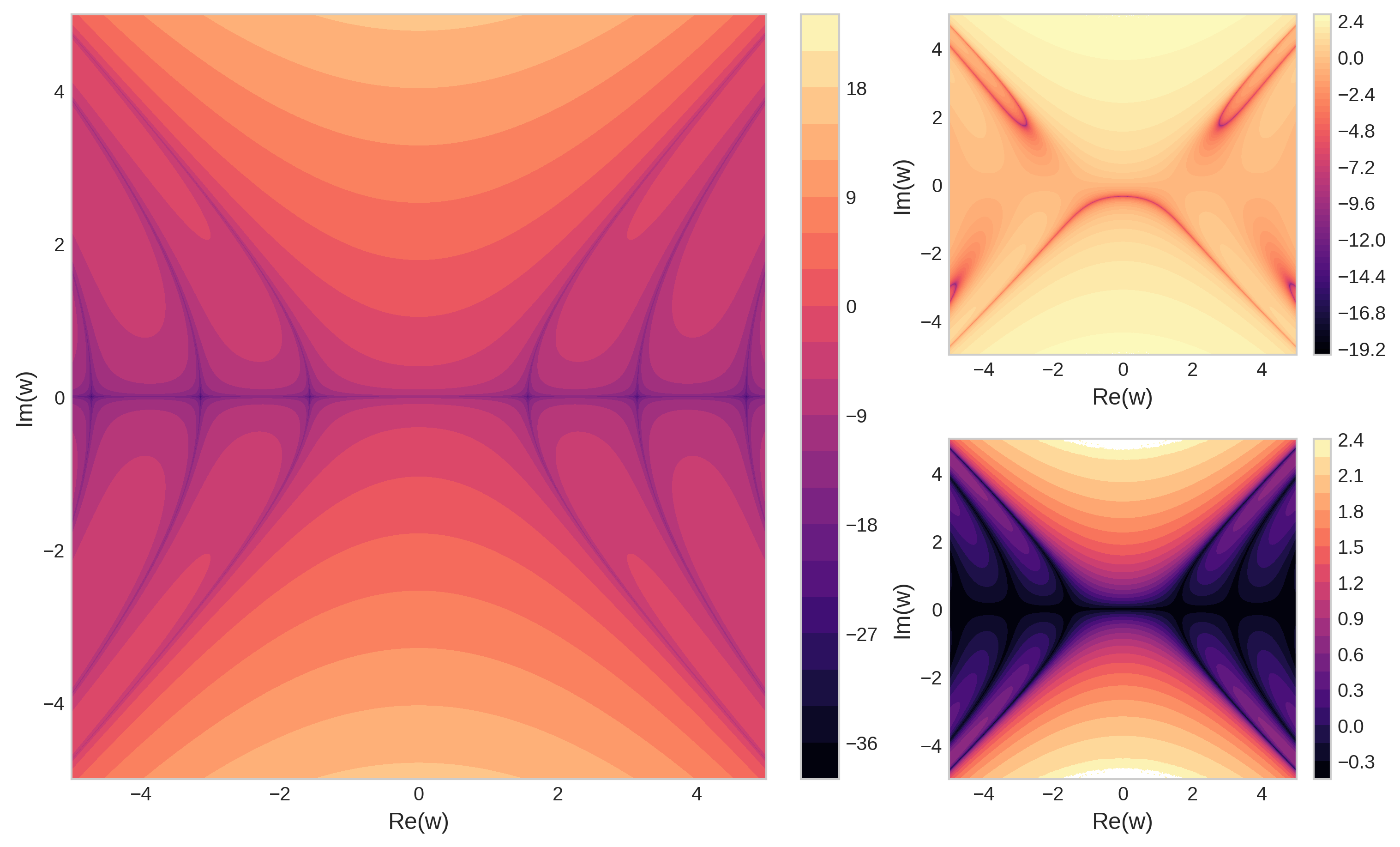}
            \caption{\label{fig:two_mode_states}Comparision of our GEM with other well-known measures of entanglement for the two-mode graph states. \textbf{Left:} Logarithm of the Geometric Gaussian Entanglement Measure. \textbf{Top Right:} Logarithm of the Entanglement of Formation. \textbf{Bottom Right:} Logarithm of the Logarithmic Negativity.}
        \end{figure*}
    
        Hence, let us introduce a family of multi-mode Gaussian states obtained by setting $c_{1,\mu\nu}=c_{4,\mu\nu}=0$ in eq. (\ref{eq:generic_preparation_hamiltonian}). We also set by convention the remaining variables to
        \begin{equation}
            c_{3,\mu\nu} = 4\,\text{Re}(A_{\mu\nu})\,,\ \ \ \ \ c_{2,\mu\nu} = 4\,\text{Im}(A_{\mu\nu})\,,
        \end{equation}
        and $A_{\mu\nu}=0$ if $\mu=\nu$.
        The states are then parameterized by the set of complex numbers $\{A_{\mu\nu}\}$ and given by
        \begin{equation}
        \label{eq:preparation_3_mode_state}
            |\psi\rangle = U\left(\left\{A_{\mu\nu}\right\}\right)|0\rangle\,,
        \end{equation}
        with
        \begin{align}
            &U\left(\left\{A_{\mu\nu}\right\}\right) = \\
            &=\text{exp}\left[i\sum_{\substack{\mu,\nu=1\\(\mu<\nu)}}^{N}\left(\text{Re}\left(A_{\mu\nu}\right)\hat{a}_\mu^\dagger \hat{a}_\nu + i\text{Im}\left(A_{\mu\nu}\right)\hat{a}_\mu \hat{a}_\nu+\text{h.c.}\right)\right]\nonumber\,.
        \end{align}
        The unitary operator generating this $M$-mode state can be expressed in quadrature basis as
        \begin{equation}
            U\left(\left\{A_{\mu\nu}\right\}\right) = \text{exp}\left[\frac{i}{2}\hat{\bm\xi}^\textsc{t}h\left(\left\{A_{\mu\nu}\right\}\right)\hat{\bm\xi}\right]\,,
        \end{equation}
        with the $h\left(\left\{A_{\mu\nu}\right\}\right)$ matrix being built out of blocks
        \begin{equation}
            \begin{pNiceMatrix}[first-row,first-col]
                & q_\nu & p_\nu \\
            q_\mu & \text{Re}(A_{\mu\nu})   & -\text{Im}(A_{\mu\nu})  \\
            p_\mu & -\text{Im}(A_{\mu\nu})   & \text{Re}(A_{\mu\nu})  \\
            \end{pNiceMatrix}\,,
        \end{equation}
        when $\mu\neq \nu$, and the trivial matrix $\bf 0_2$ if $\mu=\nu$.
        The corresponding symplectic transformation then reads:
        \begin{equation}
            S\left(\left\{A_{\mu\nu}\right\}\right)=\exp\left[\Omega h\left(\left\{A_{\mu\nu}\right\}\right)\right]\,.
        \end{equation}
        The covariance matrix is then given by
        \begin{equation}
            \Gamma\left(\left\{A_{\mu\nu}\right\}\right) = \frac{1}{2}\,S\left(\left\{A_{\mu\nu}\right\}\right) S\left(\left\{A_{\mu\nu}\right\}\right)^\textsc{t}
        \end{equation}
        The coupling constants $\left\{A_{\mu\nu}\right\}$ are naturally interpreted as the complex weights carried by the edges of a graph connecting the $N$ modes. In what follows, we will treat the case of $N=2$ and $N=3$ node graphs.
    
        \subsubsection{Two-mode states}
        \label{sec:two-mode}
    
            In the particular case of a two-mode state, we denote by $w=A_{12}$ the single coupling appearing in the Hamiltonian. The GEM can be computed exactly as a function of $w$:
            \begin{equation}
            \label{eq:two_mode_beamsplitter_gGEM}
                \text{GEM}[|w\rangle] = \frac{\text{Im}(w)^2 \sin ^2\left(2 \sqrt{\text{Re}(w)^2-\text{Im}(w)^2}\right)}{16 (\text{Re}(w)^2-\text{Im}(w)^2)}\,.
            \end{equation}
            We refer the reader to fig. (\ref{fig:two_mode_states}) (on the left) for a depiction of the GEM as a function of the complex coupling $w$. Two other well-known measures of entanglement for bipartite quantum states are depicted in the same figure: namely, the Entanglement of Formation \cite{PhysRevLett.78.5022,marian2003bures,marian2008entanglement} (computed using the algorithm of \cite{tserkis2019quantifying}) and the Logarithmic Negativity \cite {PhysRevLett.95.090503}.
            We observe that the GEM behaves qualitatively like these to other measures. Let us recall that the Entanglement of Formation is notoriously difficult to compute in general, being the solution to a complex optimization problem.
            
            By expressing the complex coupling $w$ in polar coordinates as $w = re^{i\phi}$,
            one can reexpress the GEM (\ref{eq:two_mode_beamsplitter_gGEM}) as follows:
            \begin{equation}
                \text{GEM}[|w\rangle] = \frac{\sin ^2(\phi ) \sin ^2\left(2r \sqrt{\cos (2 \phi )}\right)}{16 \cos (2 \phi )}\,.
            \end{equation}
             An interesting feature appears in the large squeezing limit. Let us set, for a moment,  $\phi=\pm\pi/2$ to turn off the pure beamsplitter component and focus on the two-mode squeezing contribution. In that case, the GEM reduces to:
             \begin{equation}
                 \text{GEM}[|w\rangle] = \frac{1}{16} \sinh ^2(2 r)\,.
             \end{equation}
             One can show \cite{serafini2017quantum} that the Schmidt decomposition (over Fock states) of the two-mode squeezed state reads:
            \begin{equation}
                \exp\left[r\left(a_1^\dagger a_2^\dagger - a_1 a_2\right)\right]|0\rangle = \frac{1}{\cosh(r)}\sum_{i\geq 0}\tanh^i(r)|i\rangle\otimes |i\rangle\,.
            \end{equation}
            In particular, in the limit of large squeezing $r\to\infty$, this state converges to the balanced superposition of all the states of the form $|i\rangle\otimes |i\rangle$. Though non-normalizable, this state, similarly to EPR states for qubits, corresponds to a maximally entangled state. Once again, this is consistent with the fact that the GEM diverges in the infinite squeezing limit.
            This observation hints towards a possible new definition of a compact GEM that is, in this case, bounded from above by the entanglement measure of such non-normalizable EPR-like states. 
            Such an upper bound of the entanglement measure can be set to $1$ by convention. For this two-mode example in particular, this can be achieved by bounding the values that can take the modulus of $w$ by substituting, for instance, $r=\tanh(\nu)$, and by normalizing the GEM as follows:
            \begin{align}
                &\text{GEM}\big[\left|\tanh(\nu)e^{i\phi}\right\rangle\big]=\nonumber\\
                &=\frac{1}{2\sinh^2(2)}\left[\frac{1}{P\left(\rho^{(1)}\right)^2} + \frac{1}{P\left(\rho^{(2)}\right)^2} - 2\right]\,.
            \end{align}
            The GEM is then bounded from above by $1$. This regularization can be viewed as effectively compactifying the space of Gaussian states by constraining the possible energy of the states \cite{serafini2007canonical, fukuda2019typical} (with respect to a Hamiltonian $\sum_\mu\tfrac{p_\mu^2+q_\mu^2}{2}$).
        
            Finally, it can be noticed that the full metric tensor can be computed exactly, as can be seen by the interested reader in the App. \ref{app:metrics}.
           
        \subsubsection{Three-mode states}
        \label{subsubsec:3mode_beamsplitter}
    
            In the case of a three-mode state, one can consider two different (connected) graph topologies. Each edge carries a complex weight; therefore,  the family of states is 6-dimensional. To visualize the GEM, we reduce the study to two separate 2-dimensional families of states.
    
            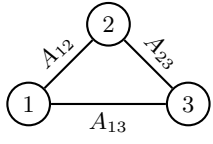
\begin{figure}[h!]
            \centering
                \begin{tikzpicture}[node distance={15mm}, thick, main/.style = {draw, circle}]
                    \node[main] (1) {$1$};
                    \node[main] (2) [above right of=1] {$2$};
                    \node[main] (3) [below right of=2] {$3$};
                    \draw (1) -- node[midway, below, sloped] {$A_{13}$} (3);
                    \draw (2) -- node[midway, above, sloped] {
                    $A_{23}$} (3);
                    \draw (1) -- node[midway, above, sloped] {$A_{12}$} (2);
                \end{tikzpicture}
            \caption{\label{fig:generic_3_node_graph}Generic complex coupling coefficients.}
            \end{figure}
    
            \paragraph*{First family:}
    
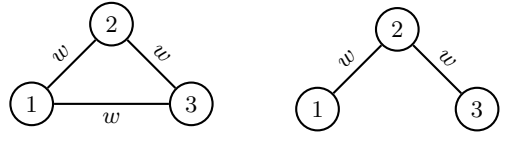
\begin{figure}[h!]
                \centering
                    \begin{tikzpicture}[node distance={15mm}, thick, main/.style = {draw, circle}]
                        \node[main] (1) {$1$};
                        \node[main] (2) [above right of=1] {$2$};
                        \node[main] (3) [below right of=2] {$3$};
                        \draw (1) -- node[midway, below, sloped] {$w$} (3);
                        \draw (2) -- node[midway, above, sloped] {
                        $w$} (3);
                        \draw (1) -- node[midway, above, sloped] {$w$} (2);
                    \end{tikzpicture}\hspace{3em}
                    \begin{tikzpicture}[node distance={15mm}, thick, main/.style = {draw, circle}]
                        \node[main] (1) {$1$};
                        \node[main] (2) [above right of=1] {$2$};
                        \node[main] (3) [below right of=2] {$3$};
                        \draw (2) -- node[midway, above, sloped] {
                        $w$} (3);
                        \draw (1) -- node[midway, above, sloped] {$w$} (2);
                    \end{tikzpicture}
                \caption{\label{fig:3_node_graph_identical_edges}First family of three-mode graph states, with identical complex couplings.}
            \end{figure}

            For the first family of states, we set all the non-zero complex couplings to be identical, and we consider both the fully connected graph (G1), and the graph with one edge turned off (G2), cf. fig. \ref{fig:3_node_graph_identical_edges}.
            
            The GEM can be expressed explicitely. For G1 we obtain:
            \begin{align}
                &\text{GEM}_{\text{G}1}\left(w\right)=\frac{\text{Im}(w)^2}{12}\frac{ \sin ^2\left(3 \sqrt{\text{Re}(w)^2-\text{Im}(w)^2}\right)}{\text{Re}(w)^2-\text{Im}(w)^2}\nonumber\\
                &=\frac{1}{12} \sin ^2(\phi ) \sec (2 \phi ) \sin ^2\left(3 r \sqrt{\cos (2 \phi )}\right)\,,
            \end{align}
            and for G2:
            \begin{widetext}
            \begin{equation}
            \begin{aligned}
                \text{GEM}_{\text{G}2}\left(w\right)&=
                \frac{\text{Im}(w)^2 \sin ^2\left(\sqrt{2} \sqrt{\text{Re}(w)^2-\text{Im}(w)^2}\right) \left(3 \cos \left(2 \sqrt{2} \sqrt{\text{Re}(w)^2-\text{Im}(w)^2}\right)+5\right)}{32 \left(\text{Re}(w)^2-\text{Im}(w)^2\right)} \\
                &=\frac{1}{32} \sin ^2(\phi ) \sec (2 \phi ) \sin ^2\left(r \sqrt{\sin (4 \phi )\csc (2 \phi )}\right) \left(3 \cos \left(2 r \sqrt{\sin (4 \phi ) \csc (2 \phi )}\right)+5\right)
            \end{aligned}
            \end{equation}
            \end{widetext}
            Note that the full metric tensor can be computed for both graphs and is provided for completeness in the App. \ref{app:metrics}. 
            
\begin{figure*}[ht]
                \centering
                \includegraphics[scale = 0.65]{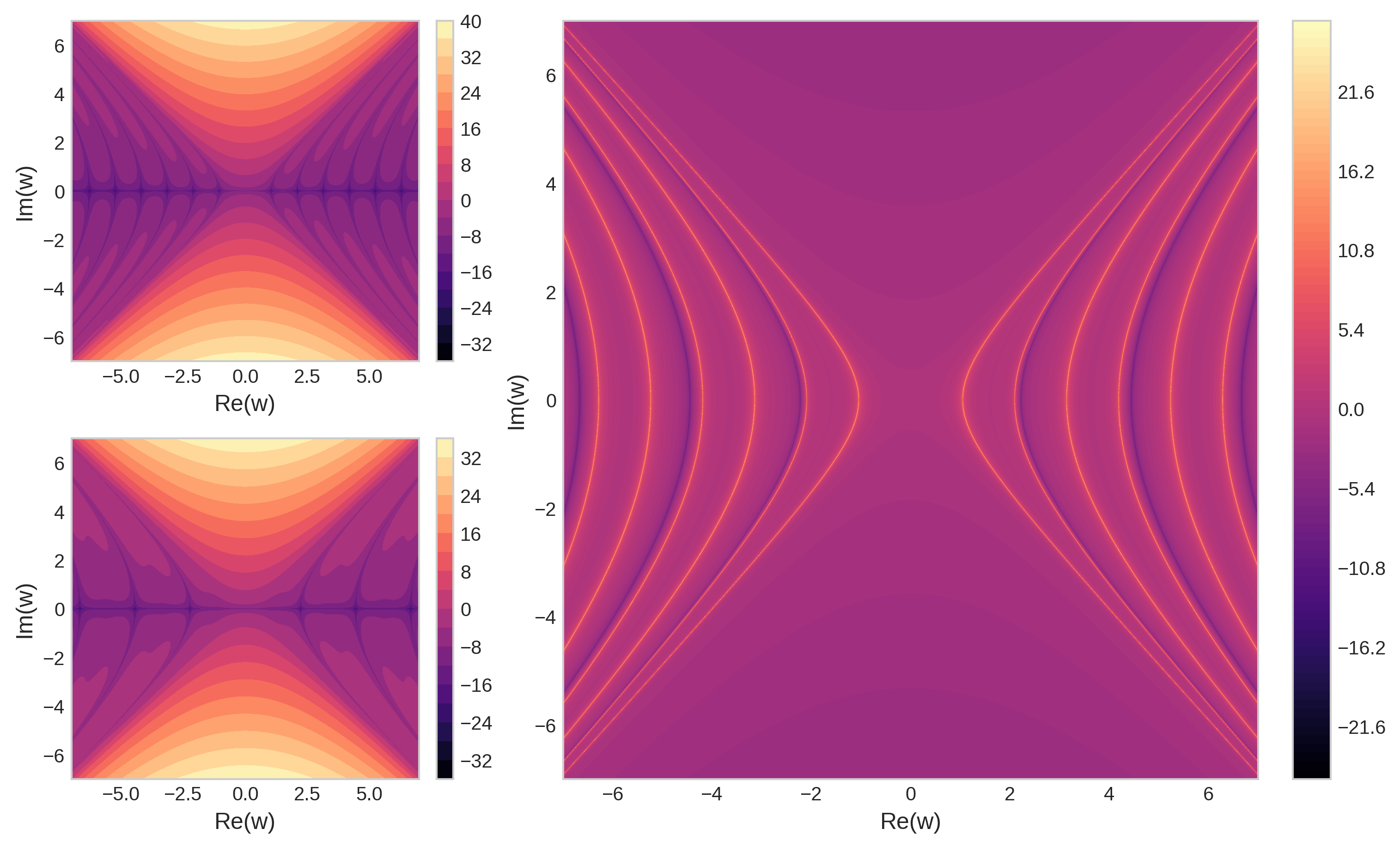}
                \caption{\label{fig:three_mode_identical_edge_weights}GEM for the first family of three-mode graph states. \textbf{Top Left:} Logarithm of the GEM for the fully connected graph. \textbf{Bottom Left:} Logarithm of the GEM for the partially connected graph. \textbf{Right:} Logarithm of the ratio of the partially connected GEM to the fully connected GEM.}
            \end{figure*}

            In fig. (\ref{fig:three_mode_identical_edge_weights}) we depict the logarithm of the GEM for these two cases, as well as the ratio of the GEM of graph 2 and graph 1. The first observation concerns the fact that a purely real coupling does not generate entanglement. To proceed with interpreting the results, it is fruitful to consider the graph weights to be all proportional to a common time parameter $t$. Therefore, an increase in the amplitude of the couplings corresponds to time evolution. Within this picture, the state preparation provided in eq. (\ref{eq:preparation_3_mode_state}) can naturally be interpreted as the unitary time evolution given by eq. (\ref{eq:preparation_3_mode_state}) of an initial state corresponding to the tensor product of Fock vacua under. Equipped with this interpretation, the comparison of the GEM for the two graph topologies becomes clear and compatible with a first intuition: the fully connected graph allows for a faster entanglement. However, for long times, the two GEM become identical, as can be seen in fig. (\ref{fig:3_node_graph_identical_edges}) (on the right), indicating that the logarithm of their ratio tends to zero for large amplitudes of $w$.
    
            Interestingly, we note that the ratio
            \begin{widetext}
                \begin{equation}
                \begin{aligned}
                    \frac{\text{GEM}_{\text{G}2}}{\text{GEM}_{\text{G}1}}(w) &= \frac{3}{8} \frac{\sin ^2\left(\sqrt{2} \sqrt{\text{Re}(w)^2-\text{Im}(w)^2}\right)}{\sin^2\left(3 \sqrt{\text{Re}(w)^2-\text{Im}(w)^2}\right)} \left(3 \cos \left(2 \sqrt{2} \sqrt{\text{Re}(w)^2-\text{Im}(w)^2}\right)+5\right) \\
                    &= \frac{3}{8}\csc ^2\left(3 r \sqrt{\cos (2 \phi )}\right) \sin ^2\left(r \sqrt{\sin (4 \phi ) \csc (2 \phi )}\right) \left(3 \cos \left(2 r \sqrt{\sin (4 \phi ) \csc (2 \phi )}\right)+5\right)\,.
                \end{aligned}
                \end{equation}
            \end{widetext}
            has the following finite limit on the two principal axes in the $w$ complex plane, corresponding to $\phi\to\frac{\pi}{4}+\frac{n\pi}{2}$ with $n\in\{0,1,2,3\}$ (or $r\to 0$):
            \begin{equation}
                \frac{\text{GEM}_{\text{G}2}}{\text{GEM}_{\text{G}1}} \to \frac{2}{3}\,,
            \end{equation}
            which is precisely the ratio of the number of edges in the two graphs.
    
            Let us mention, without entering into the details, that the above observation concerning how the ratio of GEMs captures information about the connectivity of the underlying graphs actually generalizes to more complicated graph topologies, as can be seen when inspecting the ratio of GEMs for 4-mode graph states (with obvious notations):
            \begin{equation}
            \begin{aligned}
                \frac{\text{GEM}\Big(\begin{tikzpicture}[baseline={(0,0.05)}]
                    \coordinate (A) at (0,0);
                    \coordinate (B) at (0.3,0);
                    \coordinate (C) at (0.3,0.3);
                    \coordinate (D) at (0,0.3);
                    \draw (B) -- (D);
                    \draw (A) rectangle (C);
                \end{tikzpicture}\Big)}{\text{GEM}\Big(\begin{tikzpicture}[baseline={(0,0.05)}]
                    \coordinate (A) at (0,0);
                    \coordinate (B) at (0.3,0);
                    \coordinate (C) at (0.3,0.3);
                    \coordinate (D) at (0,0.3);
                    \draw (A) -- (C);
                    \draw (B) -- (D);
                    \draw (A) rectangle (C);
                \end{tikzpicture}\Big)} \to \frac{5}{6}\,,&\ \ \ \ \ \frac{\text{GEM}\Big(\begin{tikzpicture}[baseline={(0,0.05)}]
                    \coordinate (A) at (0,0);
                    \coordinate (B) at (0.3,0);
                    \coordinate (C) at (0.3,0.3);
                    \coordinate (D) at (0,0.3);
                    \draw (A) rectangle (C);
                \end{tikzpicture}\Big)}{\text{GEM}\Big(\begin{tikzpicture}[baseline={(0,0.05)}]
                    \coordinate (A) at (0,0);
                    \coordinate (B) at (0.3,0);
                    \coordinate (C) at (0.3,0.3);
                    \coordinate (D) at (0,0.3);
                    \draw (A) -- (C);
                    \draw (B) -- (D);
                    \draw (A) rectangle (C);
                \end{tikzpicture}\Big)} \to \frac{2}{3}\,,\\
               \frac{\text{GEM}\Big(\begin{tikzpicture}[baseline={(0,0.05)}]
                    \coordinate (A) at (0,0);
                    \coordinate (B) at (0.3,0);
                    \coordinate (D) at (0,0.3);
                    \draw (A) -- (B);
                    \draw (A) -- (D);
                \end{tikzpicture}\Big)}{\text{GEM}\Big(\begin{tikzpicture}[baseline={(0,0.05)}]
                    \coordinate (A) at (0,0);
                    \coordinate (B) at (0.3,0);
                    \coordinate (C) at (0.3,0.3);
                    \coordinate (D) at (0,0.3);
                    \draw (A) -- (B);
                    \draw (B) -- (C);
                    \draw (A) -- (D);
                \end{tikzpicture}\Big)} \to \frac{2}{3}\,,& \ \ \ \ \ 
                \frac{\text{GEM}\Big(\begin{tikzpicture}[baseline={(0,0.05)}]
                    \coordinate (A) at (0,0);
                    \coordinate (B) at (0.3,0);
                    \coordinate (D) at (0,0.3);
                    \draw (A) -- (D);
                    \draw (A) -- (B);
                \end{tikzpicture}\Big)}{\text{GEM}\Big(\begin{tikzpicture}[baseline={(0,0.05)}]
                    \coordinate (A) at (0,0);
                    \coordinate (B) at (0.3,0);
                    \coordinate (C) at (0.3,0.3);
                    \coordinate (D) at (0,0.3);
                    \draw (B) -- (D);
                    \draw (A) rectangle (C);
                \end{tikzpicture}\Big)} \to \frac{2}{5}\,,\\
                \frac{\text{GEM}\Big(\begin{tikzpicture}[baseline={(0,0.05)}]
                    \coordinate (A) at (0,0);
                    \coordinate (B) at (0.3,0);
                    \coordinate (D) at (0,0.3);
                    \draw (A) -- (D);
                    \draw (A) -- (B);
                \end{tikzpicture}\Big)}{\text{GEM}\Big(\begin{tikzpicture}[baseline={(0,0.05)}]
                    \coordinate (A) at (0,0);
                    \coordinate (B) at (0.3,0);
                    \coordinate (C) at (0.3,0.3);
                    \coordinate (D) at (0,0.3);
                    \draw (A) rectangle (C);
                \end{tikzpicture}\Big)} \to \frac{1}{2}\,,&\ \ \ \ \ \frac{\text{GEM}\Big(\begin{tikzpicture}[baseline={(0,0.05)}]
                    \coordinate (A) at (0,0);
                    \coordinate (B) at (0.3,0);
                    \coordinate (C) at (0.3,0.3);
                    \coordinate (D) at (0,0.3);
                    \draw (A) rectangle (C);
                \end{tikzpicture}\Big)}{\text{GEM}\Big(\begin{tikzpicture}[baseline={(0,0.05)}]
                    \coordinate (A) at (0,0);
                    \coordinate (B) at (0.3,0);
                    \coordinate (C) at (0.3,0.3);
                    \coordinate (D) at (0,0.3);
                    \draw (B) -- (D);
                    \draw (A) rectangle (C);
                \end{tikzpicture}\Big)} \to \frac{4}{5}\,.
            \end{aligned}
            \end{equation}
    
            \paragraph*{Second family:}
    
\begin{figure}[h!]
                \centering
                    \begin{tikzpicture}[node distance={15mm}, thick, main/.style = {draw, circle}]
                        \node[main] (1) {$1$};
                        \node[main] (2) [above right of=1] {$2$};
                        \node[main] (3) [below right of=2] {$3$};
                        \draw (1) -- node[midway, below, sloped] {$1$} (3);
                        \draw (2) -- node[midway, above, sloped] {
                        $iy$} (3);
                        \draw (1) -- node[midway, above, sloped] {$ix$} (2);
                    \end{tikzpicture}\hspace{3em}
                    \begin{tikzpicture}[node distance={15mm}, thick, main/.style = {draw, circle}]
                        \node[main] (1) {$1$};
                        \node[main] (2) [above right of=1] {$2$};
                        \node[main] (3) [below right of=2] {$3$};
                        \draw (2) -- node[midway, above, sloped] {
                        $iy$} (3);
                        \draw (1) -- node[midway, above, sloped] {$ix$} (2);
                    \end{tikzpicture}
                \caption{\label{fig:3_node_graph_imaginary_edges}Second family of three-mode graph states, with  imaginary unequal couplings.}
            \end{figure}
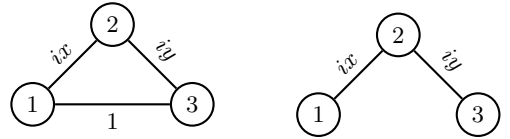

            Given that, as we saw, the real part of the coupling to not play an important role regarding the generation of entanglement in the sense of our measure, and in order to study the effect of the ratio of the communication strength between the vertices of the graph, let us now set $\text{Re}(A_{\mu\nu})=0$ and consider again the two graph topologies discussed above. For the fully connected graph (G1) we set $A_{13}=1$, and for both graphs we set $A_{12}=ix$ and $A_{23}=iy$, cf. fig. \ref{fig:3_node_graph_imaginary_edges}.
    
            Inspecting the ratio of the GEMs, we observe that asymptotically at large times (or large strength of the couplings), the ratio tends to $1$. Namely provided enough time has passed, the two states reach the same level of entanglement in the sense of our measure. However, we observe that the more balanced the strength $x$ and $y$ of the two edges are, the faster the two states reach a similar level of GEM entanglement.
    
            \begin{figure*}[ht]
                \centering
                \includegraphics[scale = 0.65]{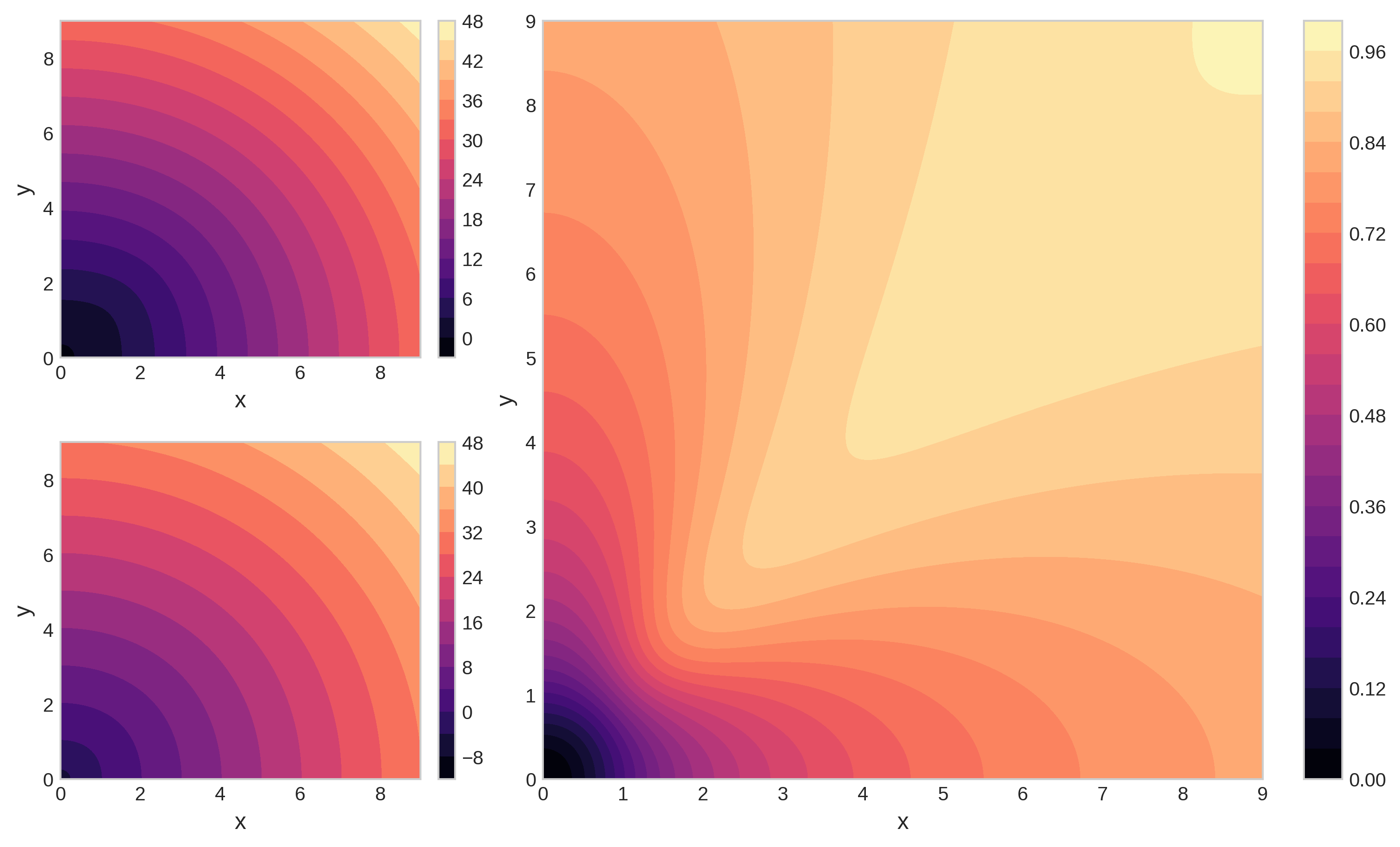}
                \caption{\label{fig:three_mode_identical_imaginary_edges}GEM for the second family of three-mode graph states. \textbf{Top Left:} Logarithm of the GEM for the fully connected graph. \textbf{Bottom Left:} Logarithm of the GEM for the partially connected graph. \textbf{Right:} Ratio of the partially connected GEM to the fully connected GEM.}
            \end{figure*}

    \subsection{Free scalar field}
    \label{subsec:scalar_field}
    
        Instead of considering systems with slightly more degrees of freedom, let us directly study the case of a very large number of bosonic modes. We therefore consider in this section a massive real Klein-Gordon field in $(1+1)$ dimensions (we set the speed of light $c$ to unity). We refer the reader to App. \ref{app:scalar_field} for details concerning this example. 
        
        We compactify the spatial dimension on a circle $S^1$ of radius $R$. The Lagrangian density of the system reads (we use the signature $(+,-)$ for the flat Lorentzian metric):
        \begin{equation}
            \mathcal L = \frac{1}{2}\Big(\partial_\mu\phi\partial^\mu\phi - m^2\phi^2\Big)\,.
        \end{equation}
        Introducing, as usual, the conjugate momentum to the field
        \begin{equation}
            \Pi = \frac{\partial\mathcal L}{\partial(\partial_0\phi)} = \partial_0\phi\,,
        \end{equation}
        the Hamiltonian density reads:
        \begin{equation}
            \mathcal H = \frac{1}{2}\Big(\Pi^2+(\bm\nabla\phi)^2+m^2\phi^2\Big)\,.
        \end{equation}
        We discretize the theory and solely consider the values of the field and its conjugate momentum on a lattice $\mathbb Z_N\subset S^1$ composed of $N=2n+1$ points $\{x_{\mu}\}_{\mu=1}^N$ separated by a distance $\delta=2\pi R/N$. The Hamiltonian of the discretized theory then reads ($Q_{N+1} = Q_1\,,P_{N+1} = P_1$)
        \begin{equation}
            \hat{H} = \sum_{\mu=1}^N\left(\frac{\delta}{2}\hat{P}_{\mu}^2+\frac{1}{2\delta}\omega^2\hat{Q}_{\mu}^2-\frac{1}{\delta^3}\hat{Q}_{\mu} \hat{Q}_{\mu+1}\right)\,,
        \end{equation}
        with
        \begin{equation}
            \hat{Q}_{\mu} = \phi(x_{\mu})\delta\,,\ \ \ \ \ \hat{P}_{\mu} = \Pi(x_{\mu})\,.
        \end{equation}
        and where we defined the effective frequency:
        \begin{equation}
            \omega = \sqrt{m^2+\frac{2}{\delta^2}}\,. 
        \end{equation}
        By translation invariance, the Hamiltonian can be diagonalized by discrete Fourier transform ($0 \leq k\leq N-1$):
        \begin{equation}
            \hat{q}_k = \frac{1}{\sqrt N}\sum_{\mu=1}^Ne^{\frac{2i\pi k\mu}{N}}\hat{Q}_{\mu}\,,\ \ \ \ \ \hat{p}_k = \frac{1}{\sqrt N}\sum_{\mu=1}^Ne^{-\frac{2i\pi k\mu}{N}}\hat{P}_{\mu}\,.
        \end{equation}
        We further separate the complex modes into real and imaginary parts by defining for all $1\leq k\leq n$
        \begin{equation}
             \begin{aligned}
            &\hat{q}_{\textsc r,k} = \frac{\hat{q}_k + \hat{q}_{-k}}{\sqrt 2}\,,\ \ \ \ \hat{p}_{\textsc r,k} = \frac{\hat{p}_k + \hat{p}_{-k}}{\sqrt 2}\,,\\
            &\hat{q}_{\textsc i,k} = \frac{\hat{q}_k - \hat{q}_{-k}}{i\sqrt 2}\,,\ \ \ \ \ \hat{p}_{\textsc i,k} = i\,\frac{\hat{p}_k - \hat{p}_{-k}}{\sqrt 2}\,,
        \end{aligned}
        \end{equation}
        in terms of which the Hamiltonian is expressed as a collection of uncoupled harmonic oscillators:
        \begin{equation}
        \begin{aligned}
            H &= \frac{\delta}{2}\,\hat{p}_0^2 + \frac{m^2}{2\delta}\,\hat{q}_0^2 +\\
            &+\sum_{k=1}^{n}\left(\frac{\delta}{2}\,\hat{p}_{\textsc r,k}^2 + \frac{\omega_k^2}{2\delta}\,\hat{q}_{\textsc r,k}^2 + \frac{\delta}{2}\,\hat{p}_{\textsc i,k}^2 + \frac{\omega_k^2}{2\delta}\,\hat{q}_{\textsc i,k}^2\right)\,,
        \end{aligned}
        \end{equation}
        with the familiar dispersion relation:
        \begin{equation}
            \omega_k = \sqrt{m^2+\frac{4}{\delta^2}\sin^2\left(\frac{\pi k}{N}\right)}\,.
        \end{equation}
        We define the dimensionless creation (and annihilation) operators by
        \begin{align}
            \hat{a}_{\mu} &= \frac{1}{\sqrt 2}\left(\sqrt{\frac{\omega}{\delta}}\,\hat{Q}_{\mu} + i\sqrt{\frac{\delta}{\omega}}\,\hat{P}_{\mu}\right)\,,\ \mu=1,\dots,N\,,\nonumber \\
            \hat{b}_0 &= \frac{1}{\sqrt 2}\left(\sqrt{\frac{m}{\delta}}\,\hat{q}_0 + i\sqrt{\frac{\delta}{m}}\,\hat{p}_0\right)\,, \\
            \hat{b}_{\textsc r, k} &= \frac{1}{\sqrt 2}\left(\sqrt{\frac{\omega_k}{\delta}}\,\hat{q}_{\textsc r, k} + i\sqrt{\frac{\delta}{\omega_k}}\,\hat{p}_{\textsc r, k}\right)\,,\ k=1,\dots,n\,.\nonumber\\
            \hat{b}_{\textsc i, k} &= \frac{1}{\sqrt 2}\left(\sqrt{\frac{\omega_k}{\delta}}\,\
            \hat{q}_{\textsc i, k} + i\sqrt{\frac{\delta}{\omega_k}}\,\hat{p}_{\textsc i, k}\right)\,,\ \ \,k=1,\dots,n\,.\nonumber
        \end{align}
        Following the conventions of \cite{blaizot1986quantum}, we gather them into vectors
        \begin{equation}
        \label{eq:index_ordering}
        \begin{aligned}
            \hat{\bm\alpha} &= \left( \hat{a}_1,\dots,\hat{a}_N,\hat{a}_1^\dagger, \dots, \hat{a}_N^\dagger\right)^{\textsc t}\,, \\
            \hat{\bm\beta} &= \left( b_0,b_{\textsc r,k},b_{\textsc i,k},b_0^\dagger, b_{\textsc r,k}^\dagger,b_{\textsc i,k}^\dagger\right)^{\textsc t}\,,
        \end{aligned}
        \end{equation}
        such that one then has $\hat{\bm\beta} = T\hat{\bm\alpha}$
        with the Bogoliubov transformation matrix:
        \begin{equation}
        \label{eq:X_Y_matrices}
        T=\left[\begin{array}{c|c}
            X^* & -Y^* \\
            \hline
            -Y & X
        \end{array}\right] \,,
        \end{equation}
        with the matrices $X$ and $Y$ given by:
            \begin{equation}
                \begin{aligned}
            X = \frac{1}{\sqrt N}&\left[\begin{array}{c}
             \frac{1}{2}\left(\sqrt{\frac{m}{\omega}} + \sqrt{\frac{\omega}{m}}\right)\bm 1_{1\times N} \\
             \left(\frac{1}{\sqrt{2}} \cos\left(\frac{2\pi k a}{N}\right)\left(\sqrt{\frac{\omega_k}{\omega}}+\sqrt{\frac{\omega}{\omega_k}}\right)\right)_{k,a}\\
            \left(\frac{1}{\sqrt{2}} \sin\left(\frac{2\pi k a}{N}\right)\left(-\sqrt{\frac{\omega_k}{\omega}}+\sqrt{\frac{\omega}{\omega_k}}\right)\right)_{k,a} 
        \end{array}\right]\,, \\
        Y = -\frac{1}{\sqrt N}&\left[\begin{array}{c}
             \frac{1}{2}\left(\sqrt{\frac{m}{\omega}} - \sqrt{\frac{\omega}{m}}\right)\bm 1_{1\times N} \\
             \left(\frac{1}{\sqrt{2}} \cos\left(\frac{2\pi k a}{N}\right)\left(\sqrt{\frac{\omega_k}{\omega}}-\sqrt{\frac{\omega}{\omega_k}}\right)\right)_{k,a}\\
            \left(\frac{1}{\sqrt{2}} \sin\left(\frac{2\pi k a}{N}\right)\left(-\sqrt{\frac{\omega_k}{\omega}}-\sqrt{\frac{\omega}{\omega_k}}\right)\right)_{k,a} 
        \end{array}\right]\,.
        \end{aligned}
            \end{equation}
        From the expressions above, one can easily check that the matrix $T$ does define a genuine symplectic transformation, namely that $T\in\text{Sp}(2N,\mathbb R)$:
        \begin{equation}
        \begin{aligned}
            XX^\dagger - YY^\dagger &= \mathbb 1\,,\ \ \ \ \ XY^{\textsc t} - YX^{\textsc t} = 0\,, \\
            X^\dagger X - Y^{\textsc t}Y^* &= \mathbb 1
            \,,\ \ \ \ \ X^{\textsc t}Y^* - Y^\dagger X = 0\,.
        \end{aligned}
        \end{equation}
        The ground state of the system $|\emptyset\rangle$ is a Gaussian state,therefore
        fully characterized by its correlation matrix. 

        Using the results reported in Appendix~\ref{subsec:der_GEM} to
        express the correlation matrix elements in terms of $X$, $Y$ matrices, the following expression for the determinant of the reduced correlation matrix:
        \begin{equation}
            \text{det}\left(\Gamma^{(\mu)}\right) =  \frac{1}{4}\left(Y^\dagger Y+X^\dagger X\right)_{\mu\mu}^2 - \dfrac{\left(Y^\dagger X\right)_{\mu\mu}^2+\left(X^\dagger Y\right)_{\mu\mu}^2}{2}\,.
        \end{equation}
        Using the expression of the matrices $X$ and $Y$, 
        one can compute this determinant very explicitly. After the dust settles down, we obtain:
        \begin{equation}
        \begin{aligned}
        \text{det}\left(\Gamma^{(\mu)}\right) = \frac{1}{4N^2}\left[1+2\sum_{k=1}^{n} \left(\frac{m}{\omega_k}+\frac{\omega_k}{m}\right)+4\sum_{k,k'}\frac{\omega_k}{\omega_{k'}}\right]\,.
        \end{aligned}
        \end{equation}
        The GEM of the ground state of the QFT then reads:
        \begin{widetext}
        \begin{equation}
        \label{eq:gGEM_field_theory}
            \text{GEM}[|\emptyset\rangle] = \frac{1}{32N}\left[1+2\sum_{k=1}^{\frac{N-1}{2}} \left(\frac{m}{\omega_k}+\frac{\omega_k}{m}\right)+4\sum_{k=1}^{\frac{N-1}{2}}\sum_{k'=1}^{\frac{N-1}{2}}\frac{\omega_k}{\omega_{k'}}\right] - \frac{N}{32}\,.
        \end{equation}
        \end{widetext}
        In the large mass limit, the GEM vanishes
        \begin{equation}
            \text{GEM}[|\emptyset\rangle] \underset{m\to\infty}{\rightarrow} 0\,,
        \end{equation}
        which is consistent with the fact that in the infinite mass limit, the coupling between neighbouring sites becomes subleading, and therefore quantum correlations are not present in the vacuum state, which simply corresponds to the (separable) tensor product of Fock vacua.
        Instead, in the limit of small mass parameter, the GEM diverges as the inverse of the mass:
         \begin{equation}
             \text{GEM}[|\emptyset\rangle] \underset{m\to 0}{\sim} \frac{1}{32\pi Rm} \cot\left(\frac{\pi}{2N}\right)\,.
        \end{equation}
        This is also consistent with the fact that the continuum theory becomes conformal in that limit because of the absence of any typical length scale.
        
\begin{figure}[ht]
            \centering
            \includegraphics[scale = 0.55]{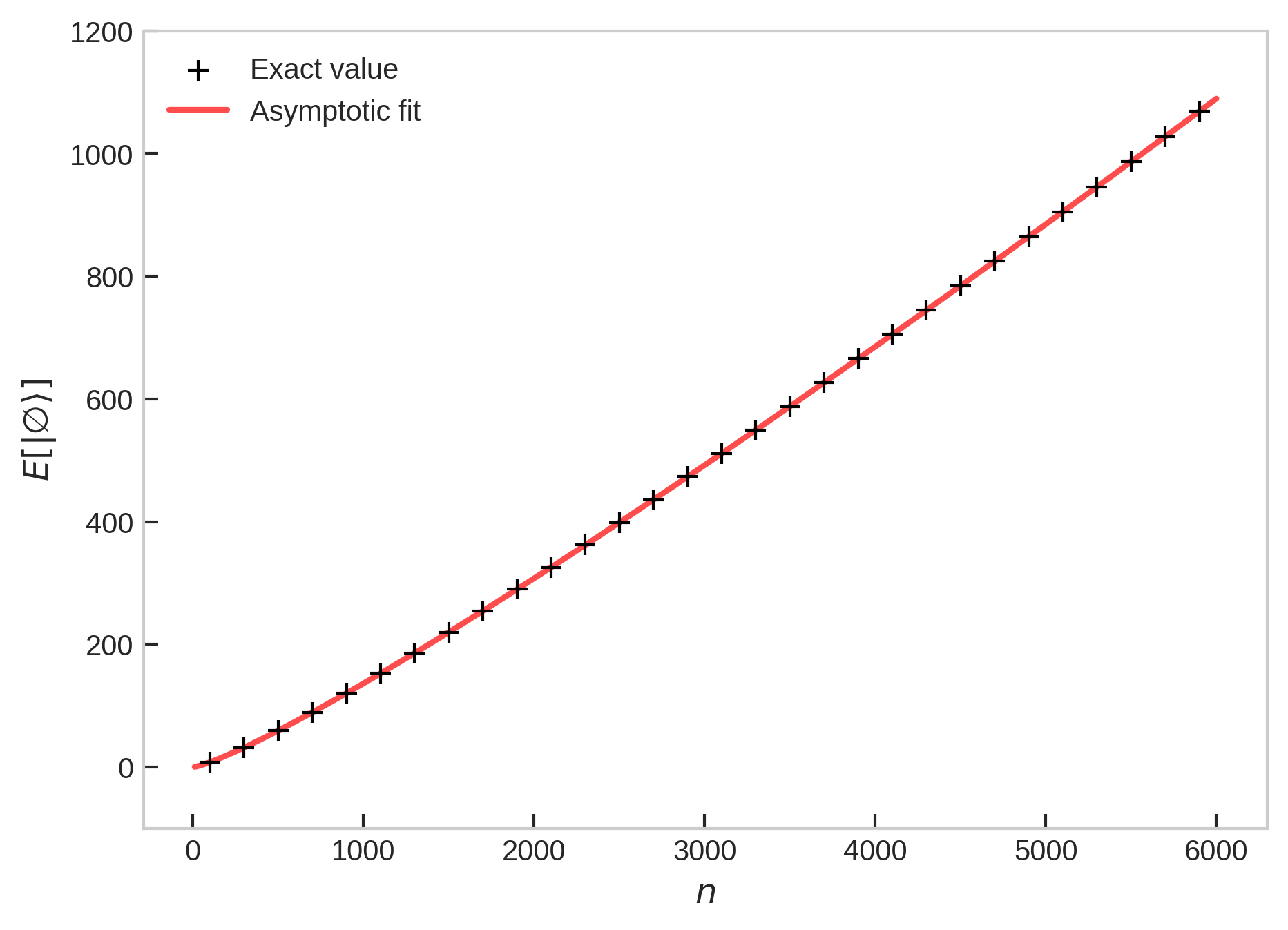}
            \caption{\label{fig:field_theory_asymptotic}Asymptotic fit of the GEM of the free massive Klein-Gordon field, with mass $m=1$ and radius of space $R=1$. We observe a perfect agreement between the exact result given by eq. (\ref{eq:gGEM_field_theory}) and the asymptotic behavior predicted by eq. (\ref{eq:field_theory_asymptotic}).}
        \end{figure}

        Another very interesting limit is the continuum limit, in which the regularization parameter $N=2n+1$ goes to infinity for fixed mass $m$ and the radius of space $R$. In App. (\ref{app:continuum_limit}), we derive the asymptotic behavior of the GEM at large $n$. We obtain the following behavior:
        \begin{widetext}
        \begin{equation}
        \label{eq:field_theory_asymptotic} 
            \text{GEM}[|\emptyset\rangle] \underset{n\to\infty}{\sim} \kappa^{(1)}_{p}(\tau) + \kappa^{(2)}_{p}(\tau)\log n + \kappa^{(3)}_{p}(\tau)\,n + \kappa^{(4)}_{p}(\tau)\,n\log n + \mathcal O(n^{-1})\,,
        \end{equation}
        \end{widetext}
        where $\tau= mR$, and where the parameter $p\in\mathbb N$, the 'Bernoulli cutoff', is explained in App. (\ref{app:continuum_limit}). Let us simply say that, in principle, the larger the cutoff, the more precise the value of the coefficients $\kappa^{(\ell)}_{p}(\tau)$, but that in practice the a priori very crude approximation $p=0$ already gives a very good estimate. 
        Therefore we observe that GEM diverges in the limit of an infinitely dense lattice in a controlled way. A few comments should be made at this point. First, note that the coefficients $\kappa^{(\ell)}_{p}$ depend on the mass and radius only through their dimensionless combination $\tau$. Second, the coefficients $\kappa^{(2)}_{p}$ and $\kappa^{(4)}_{p}$ turn out to depend neither on the Bernoulli cutoff $p$, nor on the modulus $\tau$. They are given by
        \begin{equation}
            \kappa^{(2)}_{p}(\tau) = \frac{1}{16\pi}\,,\ \ \ \ \ \ \kappa^{(4)}_{p}(\tau) = \frac{1}{4\pi^2}\,.
        \end{equation}
        Euler-Maclaurin expansions generally give good approximations already for small values of the Bernouilly cutoff, and this is indeed what we observe when plugging in some dummy values of the modulus, like $\tau=1$. We observe a stabilization of $\kappa^{(1)}_p$ and $\kappa^{(3)}_p$ already for $p=3$, and actually $p=0$ is already converGEM at two decimal figures. For concreteness, we report here the analytical expression of the running coefficients for $p=0$ and $p=1$:
        \begin{widetext}
        \begin{equation}
        \begin{aligned}
            \kappa^{(1)}_{p=0}(\tau) &= \frac{1}{32\pi}\left(\frac{1}{\sqrt{\tau^2+1}}+\frac{1}{\tau}-2 \log (\tau)-2 \log (\pi)+\log (64)+2\right)\,, \\
            \kappa^{(3)}_{p=0}(\tau) &= \frac{1}{8\pi^2}\left( \frac{1}{\sqrt{\tau^2+1}}+\frac{1}{\tau}-2 \log (\tau)-2 \log (\pi  )+\log (64)-\frac{\pi ^2}{2}\right)\,,\\
            \kappa^{(1)}_{p=1}(\tau)&=\frac{1}{192\pi}\left(\frac{1}{\left(\tau^2+1\right)^{3/2}}+\frac{6}{\sqrt{\tau^2+1}}+\frac{6}{\tau}-12 \log (\tau)-12 \log (\pi)+36 \log (2)+12\right)\,, \\
            \kappa^{(3)}_{p=1}(\tau) &= \frac{1}{28\pi^2} \left(\frac{6 \tau^2+7}{\left(\tau^2+1\right)^{3/2}}+\frac{6}{\tau}-12\log (\tau)-12 \log (\pi )+36 \log (2)-3\pi^2\right)\,.
        \end{aligned}
        \end{equation}
        \end{widetext}
        Note that one can directly check that the asymptotic behavior given in eq. (\ref{eq:field_theory_asymptotic}) is perfectly correct, as can be seen in fig. (\ref{fig:field_theory_asymptotic}).
    
        Finally, let us say that had we chosen to normalize our expression of the GEM (\ref{eq:final_expression_GEM}) with an extra factor scaling linearly in the size of the system, the above asymptotic behavior would be much simpler, and reduce to the following universal behavior:
        \begin{equation}
            \text{GEM}[|\emptyset\rangle] \underset{N\to\infty}{\sim} 
            \frac{1}{4\pi^2}\,\log N + \mathcal O\left(1\right)\,.
        \end{equation}
    
\section*{Outlook}

    In this work, we have defined the Geometric Gaussian Entanglement Measure, a simple scalar measure of quantum correlations in multimode Gaussian states. The intuition is rooted in the geometric description of the space of Gaussian states \cite{hackl2020geometry}, leveraging on the Riemannian metric and action of the local Gaussian unitary transformations, in the spirit of what was done for discrete systems in \cite{cocchiarella2020entanglement,vesperini2023unveiling}. We then computed the GEM for various natural families of Gaussian states and observed, in particular, in the multimode graph case, that it naturally captures some topological properties of the underlying graph defining the state. 

    Let us mention a few natural directions suggested by our definition. Our work naturally extends the results of \cite{cocchiarella2020entanglement} and can be viewed as a non-compact version of their qubit examples, for which the local algebra of unitaries $\bigoplus_\mu\mathfrak{su}(2)\in\mathfrak{su}(2^N)$ is naturally replaced by the symplectic group $\bigoplus_\mu\mathfrak{sp}(2,\mathbb R)\in\mathfrak{sp}(2N,\mathbb R)$. From a geometric perspective, this leads to structures reminiscent of the well-known Segre embedding for qubit systems \cite{bengtsson2017geometry}, but where the Block sphere $\mathbb{CP}^1=\text{SU}(2)/\text{U}(1)$ is now replaced by the hyperboloid $\mathbb H^1=\text{SU}(1,1)/\text{U}(1)$. More generally, the freedom in the choice of the algebra of local transformations allows for a broad generalization of our definitionon. Given an $M$-mode system, one can consider the set of partitions of the set of modes. Given a partition $P$, one associates to it the subalgebra $\bigoplus_{\sigma\in P}\mathfrak{sp}(2|\sigma|,\mathbb R)\in\mathfrak{sp}(2M,\mathbb R)$, which can then be chosen a the algebra of 'local' transformations. The choice made in this paper corresponds then to the choice of the finest partition of the set of modes. The set of partitions is naturally endowed with a partial order corresponding to the level of refinement, and we would expect our generalized GEM to respect that partial order, in the sense that given a state $|\psi\rangle$, its GEM for the coarsest partition (containing solely the whole set of modes) should vanish identically, and should be non-decreasing as the partition is being refined, reaching a maximal value for the finest partition described in this paper. This is left for future investigations. 

    We saw that the GEM can naturally be expressed in terms of a quadratic Casimir, namely a quadratic element of the center of the universal enveloping algebra.
    Though slightly deviating from the geometric root of our definition, this observation naturally suggests the possibility of defining a family of entanglement measures in terms of higher-order Casimir operators.

    Gaussian states represent an extremely rich class of continuous variable quantum states. However, including even richer families of states beyond Gaussianity is of critical importance. One possibility consists in considering the stellar representation of quantum states, which provides a neat framework in which non-Gaussianities can be implemented in a very controlled way in terms of the \textit{stellar rank} \cite{PhysRevLett.124.063605} (number of zeros of the Husimi Q-function in phase space). Another approach could be to consider families of states generated by the action of higher-order Hamiltonian, as described in \cite{guaita2021generalization}.

    Finally, it will be of great interest to consider the case of bosonic field theories in higher dimension, for instance, a free boson on $\mathbb R_t\times \Sigma_g$, where $\Sigma_g$ is a compact Riemann surface of genus $g$. Generally, the coefficients appearing in the asymptotic expansion obtained when switching off the UV cutoff of the theory could capture some invariants of the underlying manifold.

 
\section*{acknowledgments}

  The authors acknowledge funding via the FNR-CORE Grant ``BroadApp'' (FNR-CORE C20/MS/14769845) and ERC-AdG Grant ``FITMOL''.

\section*{Competing Interests}

    The Authors declare no Competing Financial or Non-Financial Interests.
    \newline


%

\newpage
\appendix
\onecolumngrid

\section{Moments}
\label{app:moments}

    We provide here some details about the moments entering in the derivation of the metric tensor in Sec. \ref{sec:derivation_metric}.

    The first moments of the $\mathfrak{sp}(2,\mathbb R)$ generators are directly given by:
    \begin{equation}
    \begin{alignedat}{2}
        M_1^\mu &= \left\langle \phi\left|\frac{q_\mu^2 - p_\mu^2}{4}\right|\phi\right\rangle &&= \frac{1}{4}\big(\mathcal C_{q_\mu, q_\mu} - \mathcal C_{p_\mu, p_\mu}\big) \\
        M_2^\mu &=- \left\langle \phi\left|\frac{p_\mu q_\mu + q_\mu p_\mu}{4}\right|\phi\right\rangle &&= -\frac{1}{4}\big(\mathcal C_{p_\mu, q_\mu} + \mathcal C_{q_\mu, p_\mu}\big) \\
        M_3^\mu &= \left\langle \phi\left|\frac{q_\mu^2 + p_\mu^2}{4}\right|\phi\right\rangle &&= \frac{1}{4}\big(\mathcal C_{q_\mu, q_\mu} + \mathcal C_{p_\mu, p_\mu}\big)
    \end{alignedat}
    \end{equation}

    For the second moments, application of Wick's theorem gives:
    \begin{equation}
    \begin{aligned}
        M_{11}^{\mu\nu}&= \frac{1}{16} \left(2 \mathcal C_{p_{\mu },p_{\nu }}^2-2 \mathcal C_{p_{\mu },q_{\nu }}^2-2 \mathcal C_{q_{\mu },p_{\nu }}^2+\left(\mathcal C_{p_{\mu },p_{\mu }}-\mathcal C_{q_{\mu },q_{\mu }}\right) \left(\mathcal C_{p_{\nu },p_{\nu }}-\mathcal C_{q_{\nu },q_{\nu }}\right)+2 \mathcal C_{q_{\mu },q_{\nu }}^2\right) \\
        M_{12}^{\mu\nu}&=\frac{1}{16} \left(4 \mathcal C_{p_{\mu },p_{\nu }} \mathcal C_{p_{\mu },q_{\nu }}-4 \mathcal C_{q_{\mu },q_{\nu }} \mathcal C_{q_{\mu },p_{\nu }}+\left(\mathcal C_{p_{\mu },p_{\mu }}-\mathcal C_{q_{\mu },q_{\mu }}\right) \left(\mathcal C_{p_{\nu },q_{\nu }}+\mathcal C_{q_{\nu },p_{\nu }}\right)\right) \\
        M_{13}^{\mu\nu} &= \frac{1}{16} \left(-2 \mathcal C_{p_{\mu },p_{\nu }}^2-2 \mathcal C_{p_{\mu },q_{\nu }}^2+2 \left(\mathcal C_{q_{\mu },p_{\nu }}^2+\mathcal C_{q_{\mu },q_{\nu }}^2\right)-\left(\mathcal C_{p_{\mu },p_{\mu }}-\mathcal C_{q_{\mu },q_{\mu }}\right) \left(\mathcal C_{p_{\nu },p_{\nu }}+\mathcal C_{q_{\nu },q_{\nu }}\right)\right) \\
        M_{21}^{\mu\nu} &= \frac{1}{16} \left(4 \mathcal C_{p_{\mu },p_{\nu }} \mathcal C_{q_{\mu },p_{\nu }}-4 \mathcal C_{q_{\mu },q_{\nu }} \mathcal C_{p_{\mu },q_{\nu }}+\left(\mathcal C_{p_{\mu },q_{\mu }}+\mathcal C_{q_{\mu },p_{\mu }}\right) \left(\mathcal C_{p_{\nu },p_{\nu }}-\mathcal C_{q_{\nu },q_{\nu }}\right)\right) \\
        M_{22}^{\mu\nu} &= \frac{1}{16} \left(4 \left(\mathcal C_{p_{\mu },q_{\nu }} \mathcal C_{q_{\mu },p_{\nu }}+\mathcal C_{p_{\mu },p_{\nu }} \mathcal C_{q_{\mu },q_{\nu }}\right)+\left(\mathcal C_{p_{\mu },q_{\mu }}+\mathcal C_{q_{\mu },p_{\mu }}\right) \left(\mathcal C_{p_{\nu },q_{\nu }}+\mathcal C_{q_{\nu },p_{\nu }}\right)\right) \\
        M_{23}^{\mu\nu} &= \frac{1}{16} \left(-4 \left(\mathcal C_{p_{\mu },p_{\nu }} \mathcal C_{q_{\mu },p_{\nu }}+\mathcal C_{q_{\mu },q_{\nu }} \mathcal C_{p_{\mu },q_{\nu }}\right)-\left(\mathcal C_{p_{\mu },q_{\mu }}+\mathcal C_{q_{\mu },p_{\mu }}\right) \left(\mathcal C_{p_{\nu },p_{\nu }}+\mathcal C_{q_{\nu },q_{\nu }}\right)\right) \\
        M_{31}^{\mu\nu} &= \frac{1}{16} \left(-2 \mathcal C_{p_{\mu },p_{\nu }}^2+2 \mathcal C_{p_{\mu },q_{\nu }}^2-2 \mathcal C_{q_{\mu },p_{\nu }}^2-\left(\mathcal C_{p_{\mu },p_{\mu }}+\mathcal C_{q_{\mu },q_{\mu }}\right) \left(\mathcal C_{p_{\nu },p_{\nu }}-\mathcal C_{q_{\nu },q_{\nu }}\right)+2 \mathcal C_{q_{\mu },q_{\nu }}^2\right) \\
        M_{32}^{\mu\nu} &= \frac{1}{16} \left(-4 \mathcal C_{p_{\mu },p_{\nu }} \mathcal C_{p_{\mu },q_{\nu }}-4 \mathcal C_{q_{\mu },q_{\nu }} \mathcal C_{q_{\mu },p_{\nu }}-\left(\mathcal C_{p_{\mu },p_{\mu }}+\mathcal C_{q_{\mu },q_{\mu }}\right) \left(\mathcal C_{p_{\nu },q_{\nu }}+\mathcal C_{q_{\nu },p_{\nu }}\right)\right) \\
        M_{33}^{\mu\nu} &= \frac{1}{16} \left(2 \mathcal C_{p_{\mu },p_{\nu }}^2+2 \mathcal C_{p_{\mu },q_{\nu }}^2+2 \left(\mathcal C_{q_{\mu },p_{\nu }}^2+\mathcal C_{q_{\mu },q_{\nu }}^2\right)+\left(\mathcal C_{p_{\mu },p_{\mu }}+\mathcal C_{q_{\mu },q_{\mu }}\right) \left(\mathcal C_{p_{\nu },p_{\nu }}+\mathcal C_{q_{\nu },q_{\nu }}\right)\right)
    \end{aligned}
    \end{equation}

\section{Free scalar field}
\label{app:scalar_field}

    In this appendix, we provide details concerning the derivation of the GEM of a free massive. 

    \subsection{Derivation of the Bogoliubov transform}
        
        As explained in Sec. \ref{subsec:scalar_field}, the starting Hamiltonian reads:
        \begin{equation}
            H = \sum_{\mu=1}^N\left(\frac{\delta}{2}P_\mu^2+\frac{m^2}{2\delta}Q_\mu^2+\frac{1}{2\delta^3}\left(Q_\mu - Q_{\mu+1}\right)^2\right) = \sum_{\mu=1}^N\left(\frac{\delta}{2}P_\mu^2+\frac{1}{2\delta}\omega^2Q_\mu^2-\frac{1}{\delta^3}Q_\mu Q_{\mu+1}\right)\,.
        \end{equation}
        where we defined the effective frequency:
        \begin{equation}
            \omega = \sqrt{m^2+\frac{2}{\delta^2}}\,. 
        \end{equation}
        The Hamiltonian can be diagonalized by discrete Fourier transform ($0 \leq k\leq N-1$)
        \begin{equation}
            q_k = \frac{1}{\sqrt N}\sum_{\mu=1}^Ne^{\frac{2i\pi k\mu}{N}}Q_\mu\,,\ \ \ \ \ p_k = \frac{1}{\sqrt N}\sum_{\mu=1}^Ne^{-\frac{2i\pi k\mu}{N}}P_\mu\,.
        \end{equation}
        We further define for all $1\leq k\leq \frac{N-1}{2}$
        \begin{equation}
            q_{\textsc r,k} = \frac{q_k + q_{-k}}{\sqrt 2}\,,\ \ \ \ \ p_{\textsc r,k} = \frac{p_k + p_{-k}}{\sqrt 2}\,,\ \ \ \ \ q_{\textsc i,k} = \frac{q_k - q_{-k}}{i\sqrt 2}\,,\ \ \ \ \ p_{\textsc i,k} = i\frac{p_k - p_{-k}}{\sqrt 2}\,,
        \end{equation}
        in terms of which the Hamiltonian reads
        \begin{equation}
            H = \frac{\delta}{2}\,p_0^2 + \frac{m^2}{2\delta}\,q_0^2 +\sum_{k=1}^{\frac{N-1}{2}}\left(\frac{\delta}{2}\,p_{\textsc r,k}^2 + \frac{\omega_k^2}{2\delta}\,q_{\textsc r,k}^2 + \frac{\delta}{2}\,p_{\textsc i,k}^2 + \frac{\omega_k^2}{2\delta}\,q_{\textsc i,k}^2\right)\,,
        \end{equation}
        with
        \begin{equation}
            \omega_k = \sqrt{m^2+\frac{4}{\delta^2}\sin^2\left(\frac{\pi k}{N}\right)}\,.
        \end{equation}
        The canonical change of variable we performed in terms of positions and momenta is therefore
        \begin{equation}
        \label{eq:canonical_transf}
            \begin{alignedat}{2}
            q_0 &= \frac{1}{\sqrt N}\sum_{\mu=1}^N Q_\mu\,, &&\ \ \ \ \ \ p_0 = \frac{1}{\sqrt N}\sum_{\mu=1}^N P_\mu\,, \\
            q_{\textsc r,k} &= \sqrt{\frac{2}{N}}\sum_{\mu=1}^N \cos\left(\frac{2\pi k\mu}{N}\right) Q_\mu\,, &&\ \ \ \ p_{\textsc r,k} = \sqrt{\frac{2}{N}}\sum_{\mu=1}^N \cos\left(\frac{2\pi k\mu}{N}\right)P_\mu\,, \\
            q_{\textsc i,k} &= \sqrt{\frac{2}{N}}\sum_{\mu=1}^N \sin\left(\frac{2\pi k\mu}{N}\right)Q_\mu\,, &&\ \ \ \ p_{\textsc i,k} = \sqrt{\frac{2}{N}}\sum_{\mu=1}^N \sin\left(\frac{2\pi k\mu}{N}\right)P_\mu\,. 
        \end{alignedat}
        \end{equation}
        We define the dimensionless creation (and annihilation) operators by
        \begin{equation}
            a_\mu = \frac{1}{\sqrt 2}\left(\sqrt{\frac{\omega}{\delta}}Q_\mu + i\sqrt{\frac{\delta}{\omega}}P_\mu\right)\,, \ \  b_0 = \frac{1}{\sqrt 2}\left(\sqrt{\frac{m}{\delta}}q_0 + i\sqrt{\frac{\delta}{m}}p_0\right)\,, \ \ b_{\textsc r/\textsc i, k} = \frac{1}{\sqrt 2}\left(\sqrt{\frac{\omega_k}{\delta}}q_{\textsc r/\textsc i, k} + i\sqrt{\frac{\delta}{\omega_k}}p_{\textsc r,\textsc i, k}\right)\,,
        \end{equation}
        in terms of which
        \begin{equation}
        \begin{aligned}
            H &= \frac{\omega}{2}\sum_{\mu=1}^N\left(a_\mu^\dagger a_\mu+a_\mu a_\mu^\dagger-\frac{1}{\delta^2\omega^2}\left(a_\mu+a_\mu^\dagger\right)\left(a_{\mu+1}+a_{\mu+1}^\dagger\right)\right) \\
            &= \frac{m}{2}\left(b_0^\dagger b_0+b_0b_0^\dagger\right) + \sum_{k=1}^{\frac{N-1}{2}}\frac{\omega_k}{2}\left(b_{\textsc r,k}^\dagger b_{\textsc r,k}+b_{\textsc r,k}b_{\textsc r,k}^\dagger + b_{\textsc i,k}^\dagger b_{\textsc i,k}+b_{\textsc i,k}b_{\textsc i,k}^\dagger\right)\,.
        \end{aligned}
        \end{equation}
        In the creation/annihilation basis, the canonical transformation (\ref{eq:canonical_transf}) reads:
        \begin{equation}
        \begin{aligned}
            b_0 &= \frac{1}{2}\frac{1}{\sqrt N}\sum_{\mu=1}^N \left[\left(\sqrt{\frac{m}{\omega}} + \sqrt{\frac{\omega}{m}}\right)a_\mu + \left(\sqrt{\frac{m}{\omega}} - \sqrt{\frac{\omega}{m}}\right)a_\mu^\dagger\right]\,, \\
            b_0^\dagger &= \frac{1}{2}\frac{1}{\sqrt N}\sum_{\mu=1}^N \left[\left(\sqrt{\frac{m}{\omega}} - \sqrt{\frac{\omega}{m}}\right)a_\mu + \left(\sqrt{\frac{m}{\omega}} + \sqrt{\frac{\omega}{m}}\right)a_\mu^\dagger\right]\,, \\
            b_{\textsc r,k} &= \frac{1}{2}\sqrt{\frac{2}{N}}\sum_{\mu=1}^N \cos\left(\frac{2\pi k\mu}{N}\right)\left[ \left(\sqrt{\frac{\omega_k}{\omega}}+\sqrt{\frac{\omega}{\omega_k}}\right)a_\mu + \left(\sqrt{\frac{\omega_k}{\omega}}-\sqrt{\frac{\omega}{\omega_k}}\right)a_\mu^\dagger\right]\,, \\
            b_{\textsc r,k}^\dagger &= \frac{1}{2}\sqrt{\frac{2}{N}}\sum_{\mu=1}^N \cos\left(\frac{2\pi k\mu}{N}\right)\left[ \left(\sqrt{\frac{\omega_k}{\omega}}-\sqrt{\frac{\omega}{\omega_k}}\right)a_\mu + \left(\sqrt{\frac{\omega_k}{\omega}}+\sqrt{\frac{\omega}{\omega_k}}\right)a_\mu^\dagger\right]\,, \\
           b_{\textsc i,k} &= \frac{1}{2}\sqrt{\frac{2}{N}}\sum_{\mu=1}^N \sin\left(\frac{2\pi k\mu}{N}\right)\left[ \left(\sqrt{\frac{\omega_k}{\omega}}+\sqrt{\frac{\omega}{\omega_k}}\right)a_\mu + \left(-\sqrt{\frac{\omega_k}{\omega}}-\sqrt{\frac{\omega}{\omega_k}}\right)a_\mu^\dagger\right]\,, \\
           b_{\textsc i,k}^\dagger &= \frac{1}{2}\sqrt{\frac{2}{N}}\sum_{\mu=1}^N \sin\left(\frac{2\pi k\mu}{N}\right)\left[ \left(\sqrt{\frac{\omega_k}{\omega}}-\sqrt{\frac{\omega}{\omega_k}}\right)a_\mu + \left(-\sqrt{\frac{\omega_k}{\omega}}+\sqrt{\frac{\omega}{\omega_k}}\right)a_\mu^\dagger\right]\,,
        \end{aligned}
        \end{equation}
        from which one reads out the matrices $X$ and $Y$ defined in eq. (\ref{eq:X_Y_matrices}).\\

    \subsection{Derivation of the GEM}
    \label{subsec:der_GEM}
    
        The ground state of the system $|\emptyset\rangle$ is a Gaussian state, therefore fully characterized by its correlation matrix.
        The correlation matrix elements in $(p,q)$-basis and $(a,a^\dagger)$-basis are related as follows:
        \begin{align}
        &\Gamma_{q_\mu,q_\mu}= \dfrac{1}{2} \langle \emptyset| (\hat{a}_{\mu}+\hat{a}^{\dagger}_{\mu})(\hat{a}_{\mu}+\hat{a}^{\dagger}_{\mu}) |\emptyset  \rangle=\dfrac{\Gamma_{a_{\mu}a_{\mu}}+\Gamma_{a_{\mu}^{\dagger}a_{\mu}^{\dagger}}}{2}+\Gamma_{a_\mu a_\mu^{\dagger}}\,,\\
        &\Gamma_{p_\mu,p_\mu}= -\dfrac{1}{2} \langle \emptyset| (\hat{a}^{\dagger}_{\mu}-\hat{a}_{\mu})(\hat{a}^{\dagger}_{\mu}-\hat{a}_{\mu}) |\emptyset  \rangle = \Gamma_{a_{\mu}a_{\mu}^{\dagger}}-\dfrac{\Gamma_{a_{\mu}a_{\mu}}+\Gamma_{a_{\mu}^{\dagger}a_{\mu}^{\dagger}}}{2}\,,\\
        &\Gamma_{p_\mu,q_\mu} = \dfrac{i}{4} \langle \emptyset| (\hat{a}^{\dagger}_{\mu}-\hat{a}_{\mu})(\hat{a}^{\dagger}_{\mu}-\hat{a}_{\mu}) +(\hat{a}^{\dagger}_{\mu}-\hat{a}_{\mu}) (\hat{a}^{\dagger}_{\mu}-\hat{a}_{\mu})|\emptyset  \rangle = \dfrac{i}{2}\left(\Gamma_{a_{\mu}^{\dagger}a_{\mu}^{\dagger}}-\Gamma_{a_{\mu}a_{\mu}}\right)\,.
        \end{align}
        It follows that the reduced determiant 
        can be rewritten as
        \begin{equation}
        \label{eq:raw_red_det}
            \mathrm{det}\left(\Gamma^{(\mu)}\right)=\Gamma_{q_\mu,q_\mu}\Gamma_{p_\mu,p_\mu}-(\Gamma_{q_\mu,p_\mu})^2= (\Gamma_{a_{\mu}a_{\mu}^{\dagger}})^2 - \dfrac{(\Gamma_{a_{\mu}^{\dagger}a_{\mu}^{\dagger}})^2 + (\Gamma_{a_{\mu}a_{\mu}})^2}{2}
        \end{equation}
        Using the inverse of the Bogoliubov matrix
        \begin{equation}
            T^{-1}=\left[\begin{array}{c|c}
                X^\textsc{t} & Y^\dagger \\
                \hline
                Y^\textsc{t} & X^\dagger
            \end{array}\right]\,,
            \end{equation}
        one can compute the correlation matrix element in $(a,a^\dagger)$-basis by exploiting the fact that the vacuum state $|\emptyset\rangle$ is annihilated by the pseudo-particle annihilation operators.
        In the $(a,a^\dagger)$-basis, one has (with index ordering prescribed by eq. (\ref{eq:index_ordering})):
        \begin{equation}
            \Gamma_{AB} = \frac{1}{2}\left\langle\emptyset\left|\left\{\alpha_A,\alpha_B\right\}\right|\emptyset\right\rangle = \frac{1}{2}\left(\mathcal C_{A,B} + \mathcal C_{B,A}\right)\,,
        \end{equation}
        with the two-point function
        \begin{equation}
            \mathcal C_{A,B} = \left\langle\emptyset\left|\alpha_A\alpha_B\right|\emptyset\right\rangle\,,
        \end{equation}
        which can be computed  In creation/annihilation basis one therefore has
        \begin{equation}
            \mathcal C_{A,B} = \left\langle\emptyset\left|\alpha_A\alpha_B\right|\emptyset\right\rangle =\sum_{C=1}^{2N}\sum_{D=1}^{2N}\left(T^{-1}\right)_{AC}\left(T^{-1}\right)_{BD}\left\langle\emptyset\left|\beta_C\beta_D\right|\emptyset\right\rangle=\sum_{c=1}^{N}\left(T^{-1}\right)_{Ac}\left(T^{-1}\right)_{B,c+N}
        \end{equation}
        and therefore
        \begin{equation}
             \Gamma_{AB} = \frac{1}{2}\sum_{c=1}^{N}\left[\left(T^{-1}\right)_{Ac}\left(T^{-1}\right)_{B,c+N}+\left(T^{-1}\right)_{Bc}\left(T^{-1}\right)_{A,c+N}\right]\,.
        \end{equation}
        One therefore obtains:
        \begin{equation}
            \text{det}\left(\Gamma^{(\mu)}\right) =  \frac{1}{4}\left(Y^\dagger Y+X^\dagger X\right)_{\mu\mu}^2 - \dfrac{\left(Y^\dagger X\right)_{\mu\mu}^2+\left(X^\dagger Y\right)_{\mu\mu}^2}{2}\,.
        \end{equation}   
        Using the expression of the matrices $X$ and $Y$, one can compute this determinant very explicitly. Indeed, one has:
        \begin{equation}
        \begin{aligned}
            &(Y^\dagger X)_{\mu\mu} = -\frac{1}{2N}\Bigg\{\frac{1}{2}\left(\frac{m}{\omega} - \frac{\omega}{m}\right)+\sum_{k=1}^{\frac{N-1}{2}}   \left(\frac{\omega_k}{\omega}-\frac{\omega}{\omega_k}\right)\Bigg\}\,, \\
            &(X^\dagger X)_{\mu\mu} = \frac{1}{N}\Bigg\{\frac{1}{4}\left(\sqrt{\frac{m}{\omega}} + \sqrt{\frac{\omega}{m}}\right)^2 +\frac{1}{2}\sum_{k=1}^{\frac{N-1}{2}} \left[\cos^2\left(\frac{2\pi k \mu}{N}\right)\left(\sqrt{\frac{\omega_k}{\omega}}+\sqrt{\frac{\omega}{\omega_k}}\right)^2 +\sin^2\left(\frac{2\pi k \mu}{N}\right)\left(-\sqrt{\frac{\omega_k}{\omega}}+\sqrt{\frac{\omega}{\omega_k}}\right)^2\right]\Bigg\}\,, \\
            &(Y^\dagger Y)_{\mu\mu} = \frac{1}{N}\Bigg\{\frac{1}{4}\left(\sqrt{\frac{m}{\omega}} - \sqrt{\frac{\omega}{m}}\right)^2 +\frac{1}{2}\sum_{k=1}^{\frac{N-1}{2}} \left[\cos^2\left(\frac{2\pi k \mu}{N}\right)\left(\sqrt{\frac{\omega_k}{\omega}}-\sqrt{\frac{\omega}{\omega_k}}\right)^2 +\sin^2\left(\frac{2\pi k \mu}{N}\right)\left(-\sqrt{\frac{\omega_k}{\omega}}-\sqrt{\frac{\omega}{\omega_k}}\right)^2\right]\Bigg\}\,,
        \end{aligned}
        \end{equation}
        and therefore
        \begin{equation}
        \begin{aligned}
            (X^\dagger X+Y^\dagger Y)_{\mu\mu} &= \frac{1}{2N}\Bigg\{\frac{1}{2}\left(\sqrt{\frac{m}{\omega}} + \sqrt{\frac{\omega}{m}}\right)^2 + \frac{1}{2}\left(\sqrt{\frac{m}{\omega}} - \sqrt{\frac{\omega}{m}}\right)^2 +\sum_{k=1}^{\frac{N-1}{2}} \left[\left(\sqrt{\frac{\omega_k}{\omega}}+\sqrt{\frac{\omega}{\omega_k}}\right)^2 +\left(\sqrt{\frac{\omega_k}{\omega}}-\sqrt{\frac{\omega}{\omega_k}}\right)^2\right]\Bigg\} \\
            & = \frac{1}{2N}\Bigg\{\frac{m}{\omega} + \frac{\omega}{m} +2\sum_{k=1}^{\frac{N-1}{2}} \left[\frac{\omega_k}{\omega}+\frac{\omega}{\omega_k}\right]\Bigg\}\,.
        \end{aligned}
        \end{equation}
        Therefore the determinant reads:
        \begin{equation}
        \begin{aligned}
            \text{det}\left(\Gamma^{(\mu)}\right) &= \frac{1}{16N^2}\Bigg\{\frac{m}{\omega} + \frac{\omega}{m}+2\sum_{k=1}^{\frac{N-1}{2}}   \left[\frac{\omega_k}{\omega}+\frac{\omega}{\omega_k}\right]\Bigg\}^2 -\frac{1}{16N^2}\Bigg\{\frac{m}{\omega} - \frac{\omega}{m} +2\sum_{k=1}^{\frac{N-1}{2}} \left[\frac{\omega_k}{\omega}-\frac{\omega}{\omega_k}\right]\Bigg\}^2 \\
            &= \frac{1}{4N^2}\left[1+2\sum_{k=1}^{\frac{N-1}{2}} \left(\frac{m}{\omega_k}+\frac{\omega_k}{m}\right)+4\sum_{k,k'}\frac{\omega_k}{\omega_{k'}}\right]\,.
        \end{aligned}
        \end{equation}
        In the limit of small mass parameter
         \begin{equation}
         \begin{aligned}
             \text{GEM}[|\emptyset\rangle] &\underset{m\to 0}{\sim} \frac{1}{32N}\left[1+\frac{4}{m\delta}\sum_{k=1}^{\frac{N-1}{2}} \sin\left(\frac{\pi k}{N}\right)+4\sum_{k=1}^{\frac{N-1}{2}}\sum_{k'=1}^{\frac{N-1}{2}}\frac{\sin\left(\frac{\pi k}{N}\right)}{\sin\left(\frac{\pi k'}{N}\right)}\right] - \frac{N}{32} \underset{m\to 0}{\sim} \frac{1}{32\pi Rm} \cot\left(\frac{\pi}{2N}\right)\,.
        \end{aligned}
        \end{equation}
    
    \subsection{Continuum limit}
    \label{app:continuum_limit}
    
        Let us write the GEM given in eq. (\ref{eq:gGEM_field_theory}) as follows (we recall that $N=2n+1$):
        \begin{equation}
            \text{GEM}[|\emptyset\rangle] = \frac{1}{32(2n+1)}\left(1+\frac{2}{m}A_n + 2m\bar A_n + 4A_n\bar A_n\right) - \frac{2n+1}{32}\,,
        \end{equation}
        with
        \begin{equation}
            A_n = \sum_{k=1}^n\omega_k\,,\ \ \ \ \ \ \ \bar A_n = \sum_{k=1}^n\omega_k^{-1}\,.
        \end{equation}
        with we recall ($R$ being the radius of space)
        \begin{equation}
            \omega_k = \sqrt{m^2 + \left(\frac{2n+1}{\pi R}\right)^2\sin^2\left(\frac{\pi k}{2n+1}\right)}\,.
        \end{equation}
        In order to obtain an asymptotic expansion for large $n$ of the two sums $A_n$ and $\bar A_n$, recall the Euler-Maclaurin formula which controls the difference between a sum and its approximating integral:
        \begin{equation}
            \sum_{k=1}^nf(k) = \int_1^n f(x)\, \text{d}x + \frac{f(1)+f(n)}{2} + \sum_{i=1}^p \frac{B_{2i}}{(2i)!}\left(f^{(2i-1)}(n) - f^{(2i-1)}(1)\right) + R_{p,n}\,,
        \end{equation}
        where the Bernoulli numbers are defined by
        \begin{equation}
            \frac{x}{e^x-1} = \sum_{i\geq 0} B_i\frac{x^i}{i!}\,,
        \end{equation}
        and where the remainder term $R_{p,n}$ is typically small. In our case we fix the integer $p$, that we will call the Bernoulli cutoff, to be small. The integral can be expressed as follows:
        \begin{equation}
            \int_1^n f(x)\, \text{d}x = n\int_0^{1-\frac{1}{n}}f(ny+1)\, \text{d}y \underset{n\to\infty}{\sim} n\int_0^1 f(ny)\, \text{d}y\,.
        \end{equation}
        Inserting for the function $f$ the functions $g(x)=\omega_x$ and $\bar g(x) = \omega_x^{-1}$, one obtains
        \begin{equation}
        \begin{aligned}
            \int_0^1 g(ny)\, \text{d}y &\underset{n\to\infty}{\sim} \frac{2m}{\pi}\,E\left(\frac{\pi}{2}\left|-\left(\frac{2n}{m\pi R}\right)^2\right.\right)\,, \\
            \int_0^1 \bar g(ny)\, \text{d}y &\underset{n\to\infty}{\sim} \frac{2}{m\pi}\,F\left(\frac{\pi}{2}\left|-\left(\frac{2n}{m\pi R}\right)^2\right.\right)\,,
        \end{aligned}
        \end{equation}
        with $E$ and $F$ being the elliptic integrals of second and first kind respectively \cite{NIST:DLMF}:
        \begin{equation}
            E(\phi|m) = \int_0^\phi\sqrt{1-m\sin^2\theta}\, \text{d}\theta\,, \ \ \ \ \ F(\phi|m) = \int_0^\phi\frac{\, \text{d}\theta}{\sqrt{1-m\sin^2\theta}}\,.
        \end{equation}
        Using these results, one can extract the asymptotic behavior of the sums $A_n$ and $\bar A_n$, and then following generic asymptotic behavior of the GEM:
        \begin{equation}
            \text{GEM}[|\emptyset\rangle] \underset{n\to\infty}{\sim} \kappa^{(1)}_{p}(mR) + \kappa^{(2)}_{p}(mR)\log n + \kappa^{(3)}_{p}(mR)\,n + \kappa^{(4)}_{p}(mR)\,n\log n + \mathcal O(n^{-1})\,.
        \end{equation}
        We refer the reader to the main text, Sec. (\ref{subsec:scalar_field}) for a discussion concerning this result.
    
\section{Metrics}
\label{app:metrics}

    For completeness, we collect here the expression of the metric tensor components for the two-mode graph states, cf. Sec. \ref{sec:two-mode}:

    \begin{equation}
        \begin{aligned}
            A &= -\frac{1}{16} \sin ^2(\phi ) \sec (2 \phi ) \sin ^2\left(2 r \sqrt{\cos (2 \phi )}\right)\,,\\
            B &= \frac{1}{16} \left(2-\sin ^2(\phi ) \sec (2 \phi ) \sin ^2\left(2 r \sqrt{\cos (2 \phi )}\right)\right)\,,\\
            C &= \frac{1}{16} \sec (2 \phi ) \left(\sin ^2(\phi ) \sin ^2\left(2 r \sqrt{\cos (2 \phi )}\right)+2 \sec (2 \phi ) \left(\cos ^2(\phi )-\sin ^2(\phi ) \cos \left(2 r \sqrt{\cos (2 \phi )}\right)\right)^2\right)\,,\\
            D &= -\frac{1}{4} \sin (\phi ) \cos (\phi ) \sec ^2(2 \phi ) \left(\cos ^2(\phi )-\sin ^2(\phi ) \cos \left(2 r \sqrt{\cos (2 \phi )}\right)\right) \sin ^2\left(r \sqrt{\cos (2 \phi )}\right)\,,\\
            E &= \frac{1}{4} \sin ^2(\phi ) \sec (2 \phi ) \sin ^2\left(r \sqrt{\cos (2 \phi )}\right) \left(\sec (2 \phi ) \sin ^2\left(r \sqrt{\cos (2 \phi )}\right)+1\right)\,.\\
        \end{aligned}
    \end{equation}
    
    \begin{equation}
    \begin{aligned}
    (h[\phi])_{(\mu,1)(\nu,1)} &= \left[
    \begin{array}{cc}
     A & B \\
     B & A \\
    \end{array}
    \right]\,, \ \ \ \ \ (h[\phi])_{(\mu,1)(\nu,2)} = \left[
    \begin{array}{cc}
     0 & 0 \\
     0 & 0 \\
    \end{array}
    \right]\,, \ \ \ (h[\phi])_{(\mu,1)(\nu,3)} = \left[
    \begin{array}{cc}
     0 & 0 \\
     0 & 0 \\
    \end{array}
    \right]\,,\\
    (h[\phi])_{(\mu,2)(\nu,2)} &= \left[
    \begin{array}{cc}
     -A & C \\
     C & -A \\
    \end{array}
    \right]\,, \ (h[\phi])_{(\mu,2)(\nu,3)} = \left[
    \begin{array}{cc}
     0 & D \\
     D & 0 \\
    \end{array}
    \right]\,, \ (h[\phi])_{(\mu,3)(\nu,3)} = \left[
    \begin{array}{cc}
     -A & E \\
     E & -A \\
    \end{array}
    \right]\,.
    \end{aligned}
    \end{equation}

\end{document}